
\input cp-aa.tex


\long\def\jumpover#1{{}}

\def\hf{\hfill}
\newdimen\thicksize
\newdimen\thinsize
\thicksize=1.8pt
\thinsize=0.6pt
\def\th{\thinspace}
\def\tth{\thinspace\thinspace}

\def\ngth{\negthinspace}

\def\hf{\hfill}

\def\frac#1#2{{\textstyle{ #1 \over #2}}}

\def\smhalf{{\scriptscriptstyle {1\over 2}}}
\def\ah   {{1\over 2}}
\def\at{{\rm\char'100}}
\def\eg{{{\sl e.g.}\ }}
\def\etal{{\sl et al.\ }}

\def\ty{{\tilde y}}
\def\tk{{\tilde k}}
\def\tomega{{\tilde \omega}}
\def\cf{{\sl cf.\ }}

\def\ie{{{\sl i.e.}\ }}

\def\viz{{\sl viz.\ }}
\def\vs{{\sl vs.\ }}

\def\ni{\noindent}

\def\Teff{{$T_{e\!f\!f} $}}
\def\Mo{{$M_\odot $}}
\def\Lo{{$L_\odot $}}
\def\dotd{\hbox{$\th.\!\!^{\rm d}$}}

\def\approxgt{\raise4pt \hbox{$>$}\kern-9pt\lower1.5pt\hbox{$\sim$}}
\def\approxlt{\raise4pt \hbox{$<$}\kern-9pt\lower1.5pt\hbox{$\sim$}}

\def\zo{z_{\rm o}}
\def\Vo{V_{\rm o}}


\overfullrule=0pt

\ni{\bf Astronomy \& Astrophysics,  in press}\vskip-10pt

\MAINTITLE{The Nature of Strange Modes in Classical Variable Stars}

\AUTHOR{
J. R. Buchler\th@1\FOOTNOTE
{e--mail address: buchler\at phys.ufl.edu\hfill},
P. A. Yecko\th@1\FOOTNOTE
{e--mail address: yecko\at phys.ufl.edu\hfill},
Z. Koll\'ath\th@1\th@2\FOOTNOTE
{e--mail address: kollath\at buda.konkoly.hu}
}
\INSTITUTE{
@1 Physics Dept., University of Florida, Gainesville FL32611, USA
@2 Konkoly Observatory, Budapest, Hungary}

\DATE{}

 \ABSTRACT{
 Strange modes have been found in the radial spectrum of many luminous stars,
such as PAGB stars.  The strange modes are characterized by having small
amplitudes in the interior of the envelope, and egregious periods and
growth-rates.  It has been common belief that the strange modes are a result of
strong nonadiabaticity.  Recently, and perhaps surprisingly, such modes have
also been found in classical Cepheid models, 
even though these are weakly
nonadiabatic stars.  Here we show that in fact there is nothing strange about
these modes and that they must exist even in the adiabatic limit.  They are
essentially acoustic surface modes.

By means of a simple change of variables and {\sl without approximation}, the
adiabatic linear pulsation equation for the radial displacement is reduced to a
Schr\"odinger like equation in which the radial coordinate is the local sound
traversal time.  In this formulation, the narrow hydrogen partial ionization
region is seen to act as a potential barrier, separating the star into two
regions.  Modes can be trapped either in the inner or in the surface region.
Coupling through the barrier gives rise to resonances between the
inner and surface regions.  The strange modes are those in which the ratio of
inner to surface amplitude is at a minimum.  The potential problem
formulation shows that strange modes exist in the adiabatic limit.  As a
function of the stellar parameters the relative location of the barrier
changes, and this gives rise to the phenomenon of {\sl avoided level crossings}
along a sequence of models.

The appearance of strange modes and the associated level crossings can be
exhibited with an {\sl analytically} solvable toy model when the potential
barrier is approximated by a delta function.  In the full nonadiabatic models
the same trapping mechanism remains responsible for the appearance of strange
modes.  The unusual growth-rates are seen to also be a consequence of the
relative minimum of the inner amplitude for these modes.  Again the behavior
of the nonadiabatic modes can be well mimicked by a simple analytical toy
model.

The strange modes can be linearly unstable to the left of the fundamental and
first overtone blue edges.  Hydrodynamical calculations show that the strange
limit cycle pulsations (a) are extremely superficial as the linear eigenvectors
already indicate, in fact they have negligible amplitudes interior to the
partial hydrogen ionization front, and \th (b) the pulsations have surface
radial velocities in the 0.1 -- 1.0 km/s range, but extremely small
photospheric velocities, and luminosity variations in the milli-magnitude
range. \th These modes are therefore expected to be difficult to observe.  }


\maketitle

\input psfig

\def\rahmen#1{}

\titlea{Introduction}

This work sets out to demystify the so-called strange vibrational modes that
are regularly encountered in stars with high luminosity to mass ratio, such as
PAGB stars (Cox et al. 1980, Saio et al. 1984, Aikawa 1985, Gautschy \& Glatzel
1990, Zalewski 1992, 1994, Cox \& Guzik 1996, Papaloizou et al. 1997), 
but more recently also in
Cepheids (Buchler 1997) where they were not expected because of the weakly
nonadiabatic nature of the pulsations in these stars.  We begin in \S2 by
rephrasing the description of adiabatic pulsations in the general form of a
quantum mechanical potential well problem.  The natural follow-up is to focus
on the resulting potential.  This we do for a Cepheid and find that the
sharpness of the hydrogen ionization front produces an immensely high barrier
in the potential, effectively separating the star into two distinct regions.
The barrier is ultimately caused by the rapid variation of the local sound
speed, most by the adiabatic index.  In \S3, for didactic purposes,
we present a toy problem that approximates the Cepheid, yet contains all the
essential features of strange modes, and that can be solved analytically.  We
thereby verify that the first appearance of a strange mode occurs in concert
with a node crossing the potential barrier into the narrow outer envelope of
the star between the surface and the hydrogen partial ionization region (HPIR).

In \S4 we look in detail at a specific Cepheid model, using the transformations
inspired by the potential well formulation set out in \S2.  An entire sequence
of Cepheid models is then examined in \S5, and the signature of strange modes
is revealed unequivocally.  It will become apparent that the gradual change in
the model parameter, which affects the extent of the region between the HPIR
and the stellar surface, modifies the resonance conditions between the inner
and outer regions and is responsible for the observed systematic changes along
the sequence.  In \S6 we extend the adiabatic toy model of \S3 to the
nonadiabatic case by allowing a complex potential barrier.  Although we show
this model to be highly approximate, it does reproduce the behavior of the
periods and of the growth-rates in Cepheid sequences quite accurately.

The pulsating stellar envelope, both physically and mathematically, bears a
strong resemblance to a (quarter wave length) flared
wind instrument such as a horn or
trumpet.  In \S7 we explore this analogy and design an {\sl equivalent musical
wind instrument} with the characteristics of the Cepheid spectrum.

The strange modes can be unstable outside the usual Cepheid instability strip
and can thus give rise to nonlinear pulsations of their own.  In order to shed
some light on the appearance of such pulsations and answer the question of
their observability we therefore compute the nonlinear hydrodynamical behavior
of several such ``Cepheid'' models in \S8.

 \begfig0cm
 \centerline{\psfig{figure=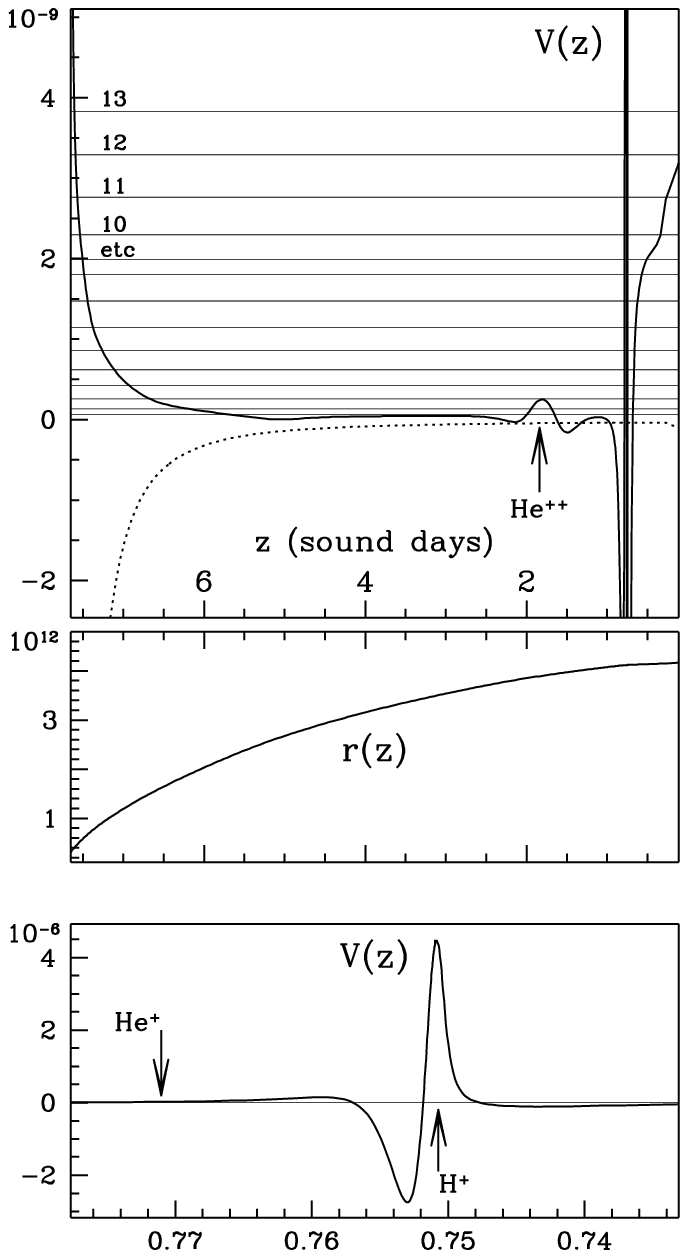,width=8.5cm}}
 \vskip -10pt
 \figure {1}{Effective potential for the adiabatic pulsations;
 {\sl top}: full potential (solid line) and gravitational contribution (dotted
line);  the horizontal lines denote the lowest eigenvalues $\omega_n^2$;
 {\sl middle:} scale transformation $r(z)$;
 {\sl bottom}: blow-up of the region near the hydrogen partial ionization
region;
 Stellar surface is on the right at $z=1$.
 }
 \endfig

\titlea{The Cepheid Potential}

In the standard equilibrium radiative diffusion approximation adiabatic radial
pulsations are described by the familiar equation for the radial displacement
eigenfunctions $\delta r$:
 $$
{\partial^2\over\partial t^2} (r^2\delta r) =
 {r^2\over \rho}{\partial\over\partial r}\left[
 {\Gamma_1 P\over r^2}{\partial\over\partial r} (r^2 \delta r)\right]
 + {4GM\over r^3}r^2\delta r\; .
 \eqno(1)$$
 where the symbols have their usual meanings (\eg Cox 1980).

 We can recast the description of the adiabatic pulsation
problem in terms of a new function $\Phi$ by invoking the transformation
 $$
 \Phi = \left(\Gamma_1 P\right)^{\smhalf} r \delta r 
 = \left(\rho\right)^{\smhalf} c_s r \delta r \; ,
 \eqno(2)$$
 where the second equality simply involves 
the adiabatic sound speed: $c_s^2 = \Gamma_1 P/\rho$.
 Thus
 $$
 c_s^2\tth {d^2 \Phi\over dr^2} +W(r)\Phi
 = \omega^2 \Phi \; ,
 \eqno(3)$$
  where the assumed $exp(-i\omega t)$ time dependence has allowed us
to replace the temporal derivative by the eigenvalue $\omega^2$,
and the function $W(r)$ is given by
 $$
 W(r) = {1\over \left(\rho^{\smhalf} c_s/r\right) } 
 {d^2\over dr^2} \left(\rho c_s^2 / r^2\right)^{\smhalf} 
 -{4GM\over r^3} \; .
 \eqno(4)$$
 The first term of $W(r)$ embodies the effects of the spatially varying inertia
(density) and sound speed on $\Phi$, and the form it takes is a result of
applying Eq.~(2) to Eq.~(1).  Anywhere the equilibrium varies rapidly, $W(r)$
is dominated by this first term.  The second term of $W(r)$ results from the
variation of gravitational forces within the hydrostatic equilibrium and will
only be important at small radius and for modes with small wave number (these
have local wavelengths comparable to the scale for which the gravitation
varies).  Both terms play a role in determining the global frequency spectrum
of radial modes.

We note in passing that when a plane-wave function $\Phi(r)=\exp (ikr)$ is
inserted, Eq.~(3) then represents the local dispersion relation and can be used
to construct a {\sl propagation diagram} (Unno et al. 1989).

We can now go one step further and convert Eq.~(3) into a Schr\"odinger like
equation by eliminating the sound speed that modifies the spatial
derivative in Eq.~(3); we do this by transforming from $r$ to a new
coordinate $z$, chosen to increase inward:
 $$
 dz = -dr/c_s.
 \eqno(5)
 $$
 The quantity $z$ then represents the {\sl acoustic depth} (or sound
traversal time) and is conveniently measured in {\sl sound days} here
($r(z)$ appears in Fig.~1, center panel).  The newly phrased problem
involves a concomitant transformation of the eigenfunctions according
to:
 $$
 \Psi  =  (\rho c_s)^{\smhalf} r \delta r = c_s^{-\smhalf} \Phi
 \eqno(6)$$
 which converts Eq.~(1) into
 $$
 -{d^2 \over dz^2} \Psi + V(z)\th \Psi
 = \omega^2 \Psi \; ,
 \eqno(7)$$
 where the potential is given by
 $$
 V(z) =
 {1\over \left(\rho c_s / r^2\right)^{\smhalf} } 
 \tth {d^2\over dz^2} \left(\rho c_s / r^2\right)^{\smhalf}
 -{4GM\over r^3}   \; .
 \eqno(8)$$ 
 Note that the potential $V$ in Eq.~7 differs from $W$.

 Eq.~(7) is clearly the simplest and most compact form that the problem of
adiabatic stellar pulsations can be expressed in.  The spatial gradients of the
equilibrium quantities $\rho$ and $c_s$, as well as sphericity effects have
been incorporated into the modified eigenfunctions $\Psi$ and into the
transformed spatial coordinate $z$.  The effective sound speed is therefore
unity and we can set $\omega = k$, the wave-vector.  The boundary conditions
associated with the original problem Eq.~(1) are readily transcribed to the new
formulation of Eq.~(7).

 Eq.~(7) has been derived for the radial displacement eigenvector, or
equivalently the velocity.  In the literature one also finds equations similar
to Eq.~(3) or (7), but for the {\sl pressure} eigenfunction $\delta p$ (\eg
Deubner \& Gough 1984).  We stress that the derivation for such a Schr\"odinger
equation for $\delta p$ is only possible if approximations are made.  The exact
equation for $\delta p$ would involve third order spatial derivatives, unless
the variations of the gravitational potential are neglected in the
linearization (Cowling approximation).

\vskip 10pt

 We now examine the potential of a specific Cepheid model, choosing the
parameters $L=3000$\Lo, $M=5$\Mo, \Teff=5520\th K, $Z=0.02$, $X=0.70$.  We use
the latest OPAL opacities (Iglesias \& Rogers 1996) with Alexander-Ferguson
(1994) molecular opacities.  Convection is ignored.

 The top panel of Fig.~1 displays $V(z)$ (heavy solid line) and its
gravitational component (dotted line) for this Cepheid model.  The gravitation
contributes a small, but significant depth in the interior.  In this plot the
coordinate $z$ ranges over the entire pulsating envelope.  The scale of $z$ is
of the same order as the period of the gravest pulsational modes (cf. virial
theorem).  In order to reveal the overall structure of the potential well
the vertical scale is highly magnified.  The horizontal lines indicate
the relative placement of the adiabatic pulsation energies.  The barrier at
$z=0\dotd 75$ is so high as to be off-scale and so narrow as to be featureless
with respect to the scales that are appropriate for the pulsations.  Therefore
in the bottom panel of Fig.~1 we exhibit the shape of the potential $V(z)$,
restricting our perspective here to a very narrow region in the vicinity of the
HPIR.

 \begfig0cm
 \centerline{\psfig{figure=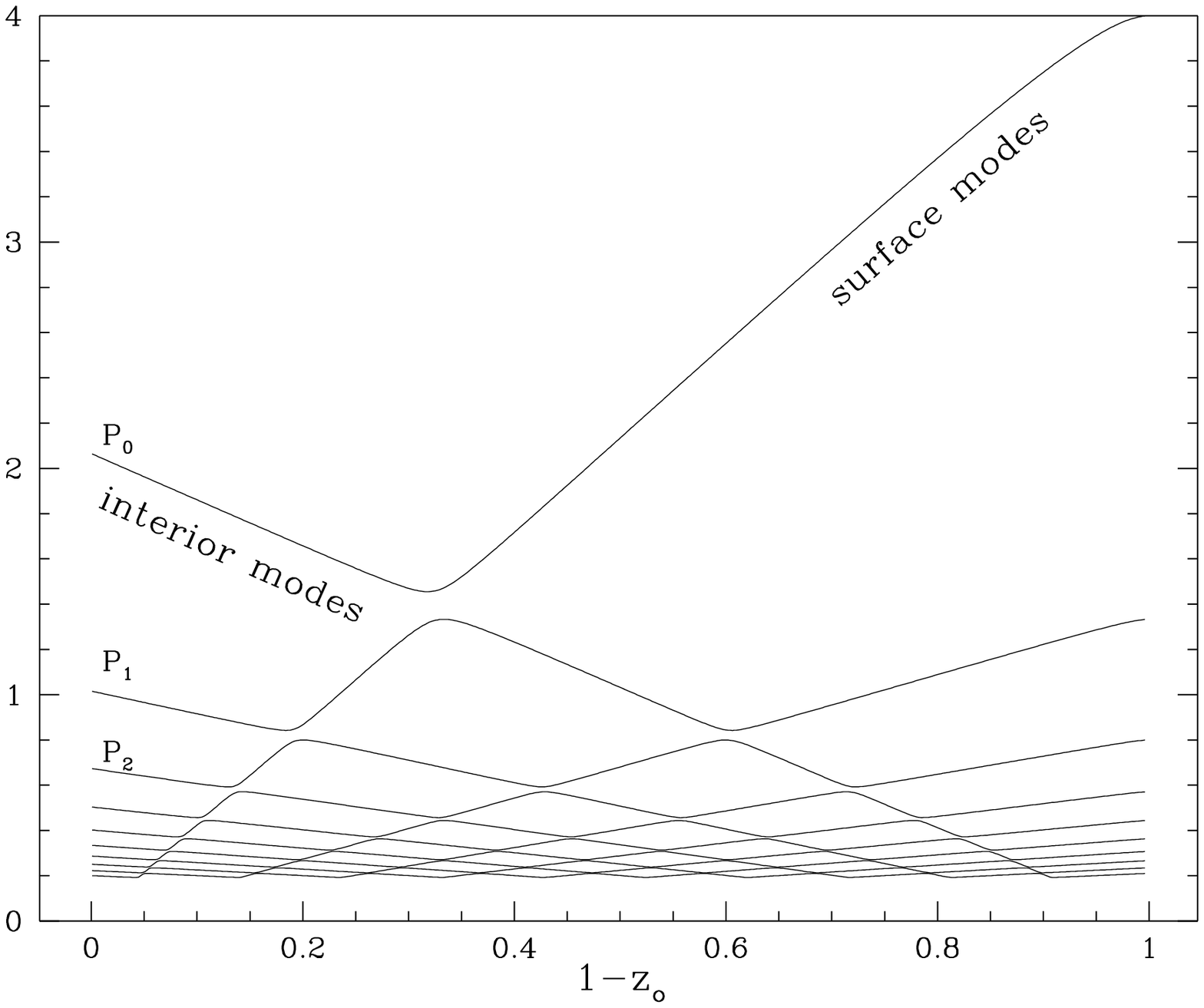,width=9.cm}} 
 \vskip -10pt
 \figure {2a}{Periods of the barrier problem, with $\Vo$=10; $\zo$ is the
distance of the barrier from the inner (nodal) boundary; $1$-$\zo$ is thus
the width of the outer region.
 }
 \centerline{\psfig{figure=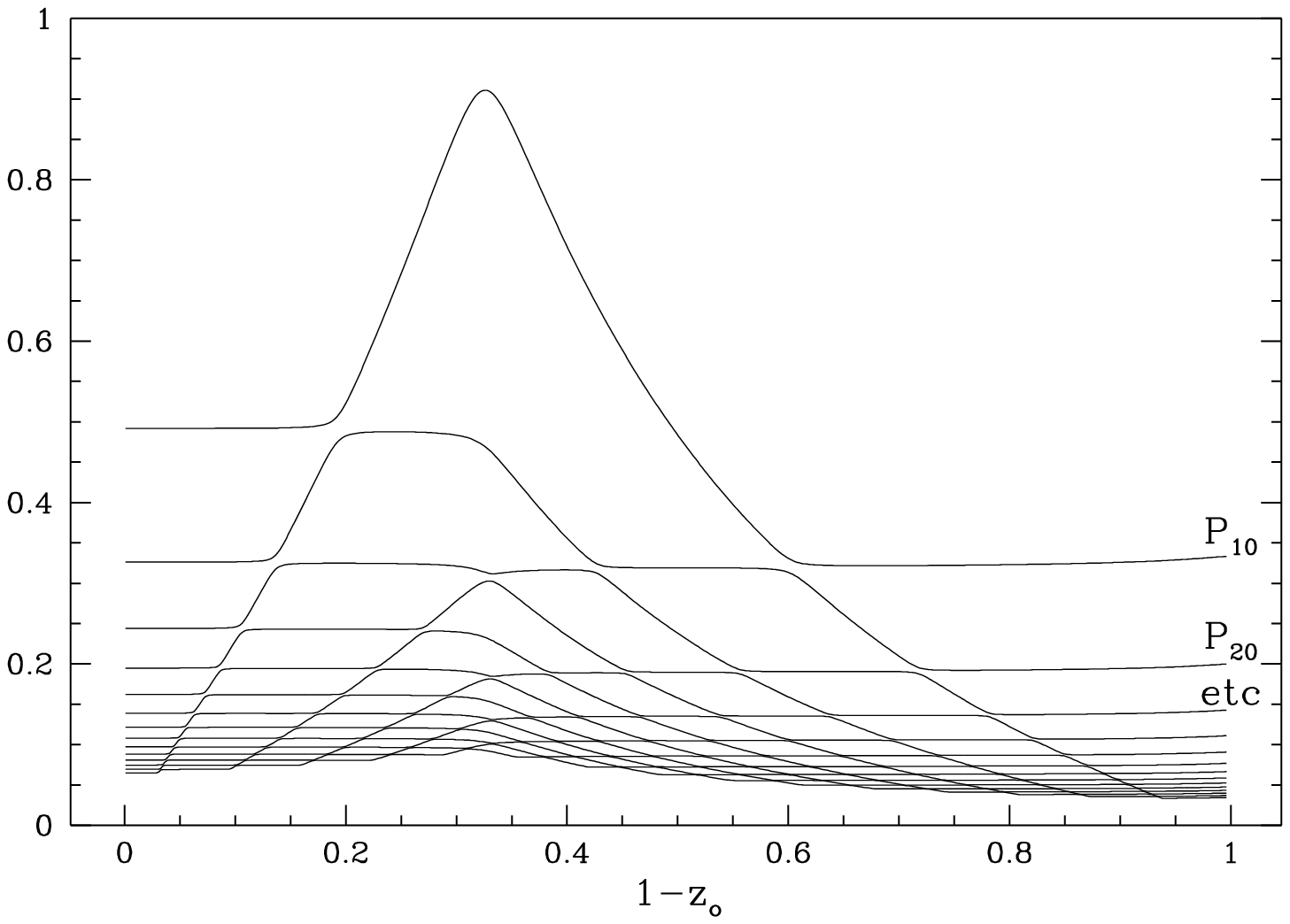,width=9.cm}}
 \vskip -10pt
 \figure{2b}{Period ratios of the barrier problem, with $\Vo$=10; $1$-$\zo$ is
 the width of the outer region.
 }
 \endfig

 \begfig0cm
 \centerline{\psfig{figure=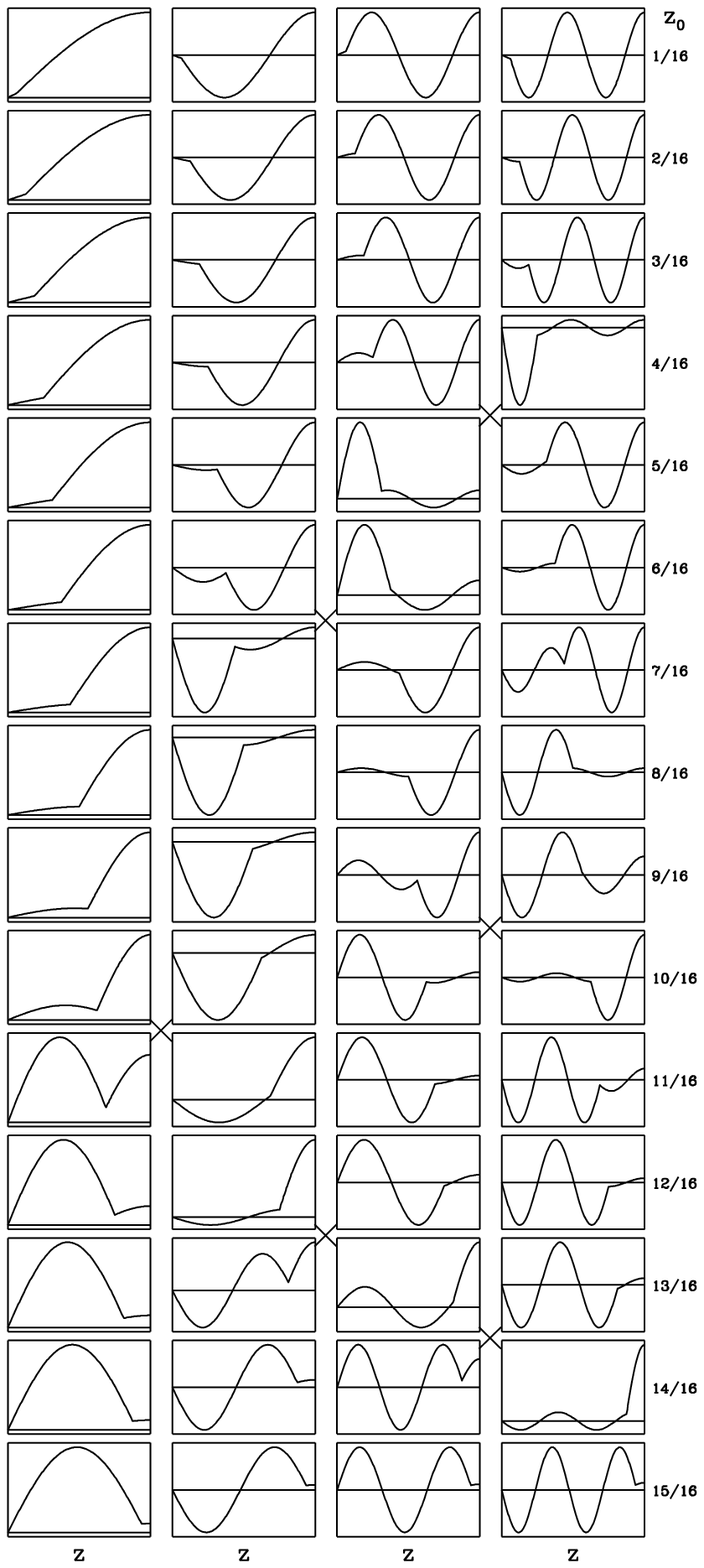,width=9.cm}} 
 \vskip -10pt
 \figure {3}{Lowest four eigenfunctions as a function of barrier location
$\zo$, with $\Vo$=10; $1$-$\zo$ is the width of the outer region.
 }
 \endfig

The middle panel of Fig.~1 plots the relation between the radial coordinate $r$
and the acoustic depth $z$.  In the new $z$ coordinate the inner regions with
their high sound speed now appear highly compressed.  We have adopted zero
acoustic depth ($z=0$) at the stellar surface.

Of the three quantities that appear in the second derivative of Eq.~8, it is
the sound speed whose variations are most significant.  Recall that $c_s
\propto \Gamma^\ah$ and $\Gamma$ experiences rapid dips (\eg Cox \& Giuli 1968)
in ionization zones that then produce the large potential fluctuations.  The
centers of the
three ionization stages, H--H$^+$, He--He$^+$, He$^+$--He$^{++}$ are indicated
by arrows in Fig.~1.  Of these it is the hydrogen that is by far the narrowest,
leading to a potential barrier.

\titlea{An Idealized Potential Barrier Toy Problem}

For a potential like that shown in Fig.~1 the eigensolutions have many
properties that can be demonstrated in a more transparent way by looking at a
toy problem.  This toy problem can be solved analytically, yet it displays the
essential features of the modal spectrum of the star.  We replace the actual
potential with an infinitesimally narrow barrier (a delta function 
of strength $\Vo$) which is
located at $\zo$ \tth ($0<\zo<1$) and the potential 
simply zero elsewhere.  Thus the Schr\"odinger
equation reduces to
 $$-y'' + \Vo\th k\delta (z-\zo)\th y = \omega^2 y .  \eqno(9)
 $$ 
 where $k\delta (z-\zo)$ models the potential shape and $\Vo$ is a
dimensionless scale factor.  We choose the boundary conditions appropriate for
the adiabatic pulsations of a Cepheid, those for a quarter wave-length tube
that is closed at the inner boundary and open at the outer (pure reflection),
\viz :
 $$y(0) = y'(1) = 0 \; .
 \eqno(10)$$
 The eigenvectors are given by
 $$\vcenter{\openup1\jot
   \halign{
 $\hfil#$ & ${}#\hfil$ &
  \qquad \qquad \qquad\quad \qquad $\hfil#$ & ${}#\hfil$   \cr
 y(z) &= a\tth  \sin kz      &    z \leq \zo  \cr
      &= \cos k(1-z)        &    \zo \leq z  \cr}
 }\eqno(11)$$
 with normalization $y(1)=1$ \th(more generally $a(k)=y'(0)/ky(1)$).  
With the appropriate matching conditions 
 $$
 \eqalignno{         
 y(\zo^+)  &= y(\zo^-)                      &(12a)\cr
 y'(\zo^+) &= y'(\zo^-) +k\th\Vo y(\zo)\,,    &(12b)\cr
 }
 $$
 one obtains the characteristic equation for $k$
 $$\cos k + \Vo \th \sin  k \zo \th \cos  k (1-\zo)=0 \;.
 \eqno(13)$$
 The inner amplitude $a$ is
given by
 $$a(k)  = \cos  k (1-\zo)/\sin k \zo .
 \eqno(14)
 $$

 \begfig0cm
 \centerline{\psfig{figure=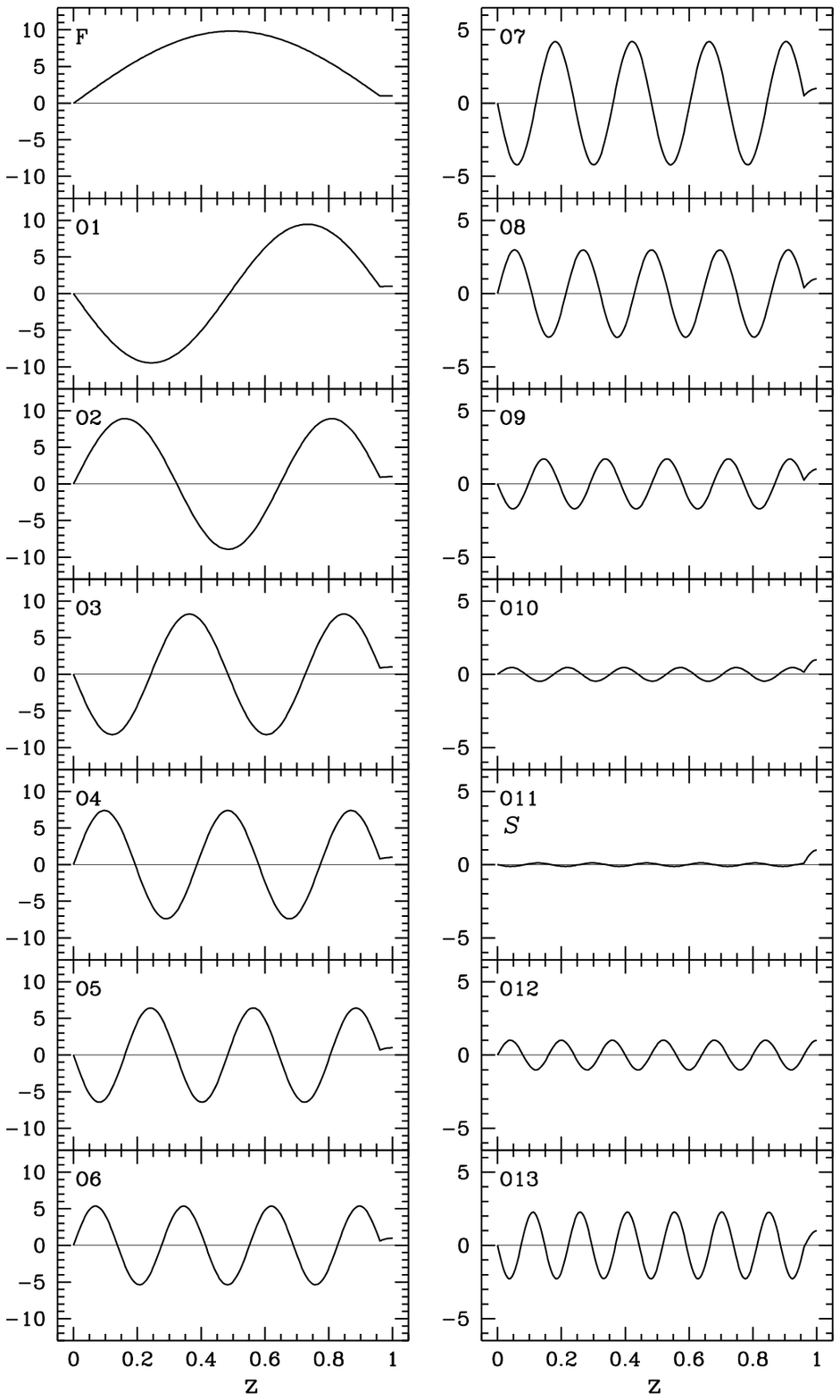,width=9.cm}}
 \vskip -10pt
 \figure {4}{Eigenvectors $y_n$ of the potential barrier toy problem with
$\zo=0.96$ and with $\Vo$=10; the strange mode is labelled ${\cal S}$.
 }
 \endfig

We note several interesting features.  First, Eq.~13 shows that for large $\Vo$
(a very high barrier) one obtains $\cos k (1\ngth-\ngth\zo) \sin k \zo \approx
0$.  Thus for $\zo\rightarrow 0$ (vanishing inner region) the spectrum is given
by $\cos k=0$, with periods $P_n=4/(2n+1)$, i.e. the one of an open tube,
whereas for $\zo \rightarrow 1$ (vanishing outer region) we have essentially a
closed tube and its spectrum $\sin k =0$ with $P_n\approx 2/(n+1)$.  This
asymptotic behavior in $\zo$ is nicely seen in Fig.~2a that shows the period
spectrum as a function of 1--$\zo$ for a value $\Vo$=10.  (The reason for
plotting versus 1-$\zo$ instead of $\zo$ 
is that we eventually compare to fig.~10 in which the outer
acoustic distance increases with $P_0$.)  Following common
practice we assume that the modes are ordered by decreasing periods $P_n$.

 A null inner amplitude can occur for specific combinations of $\Vo$ and $\zo$,
\ie nongenerically, as Eq.~(14) shows.  We call these modes {\sl perfect}
surface modes.  It will be seen later in the application to stars that the
surface modes are in fact the strange modes.  For arbitrary values of $\Vo$ and
$\zo$ we refer more loosely to the mode whose amplitude is at a minimum as the
strange mode.

 The outer boundary condition precludes a zero outer amplitude, i.e. infinite
values of $a$, but $a$ can get very large.  We refer to the corresponding modes
as {\sl inner} modes; later we shall see that in stars the inner modes are the
{\sl regular} modes.

 Finally, in between the inner and the outer modes lie modes that have a
uniform amplitude throughout ($a$=1).  Such perfectly matched modes can occur
accidentally when a node happens exactly at $\zo$ (\cf Eq.~12b).  For most
values of $\Vo$ and $\zo$ one can generally identify modes that are close to
being matched.

It is amusing to notice in Figs.~2aand 2b how, 
through the intermediary of trapped
modes, the spectrum gradually changes from that of a closed to a open tube as
$1$-$\zo$ varies between 0 and 1.  The reason for this, as we have seen, is
that a very strong barrier (large $V_0$) acts almost like an additional interior
boundary condition.  Below the barrier, the modes resemble those of a closed
pipe of length $\zo$, while above the barrier the modes resemble those of a
open pipe of length $1$-$\zo$.  Except for the rare situation of matched modes,
either the inner or the outer amplitude dominates, i.e. the modes are either
inner (regular) or outer (strange) modes.

The appearance and behavior of the modes is further, and more globally
illustrated in Fig.~3 which shows the lowest three eigenvectors as a function
of $z$.  We recall that the closed end is on the left ($z=0$) and the open end
on the right ($z=1$).  In the right margin we show the location of the barrier
$\zo$.  This figure should be looked at in conjunction with Fig.~2a which shows
the periods $P_n$ as a function of $\zo$.

Consider first the fundamental mode displayed in the leftmost column of Fig.~3.
As expected, for small $\zo$ the outer well dominates and the eigenfunction is
essentially that of an open, quarter-wavelength tube.  As the barrier moves to
the right, the effective length of the tube shrinks, causing a decrease in the
period as seen in Fig.~2a.  On the other hand, starting from the bottom, we see
that the eigenfunction is now essentially that of a closed half wavelength tube
with half the period of the $\zo=0$ case.  For $\zo$=2/3 one can fit exactly
3/4 of a wavelength into the well, which corresponds to a perfect match
condition.  At the shown values of $\zo$=10/16 and 11/16 the $y_0$ eigenvectors
are indeed very close to those of the first overtone of an open tube, but being
the fundamental mode they avoid acquiring a node.  This is also the place where
a near degeneracy occurs as Fig.~2a indicates.  However, because the
Sturm-Liouville nature of the problem precludes degeneracy, there must be an
avoided level crossing. (Strictly speaking, the problem is not Sturm-Liouville
because of the $\delta$ function potential, but seems to preserve the
features thereof in this singular limit).  If we look at the second column
which represents the first overtone in the well we notice that the system
avoids degeneracy through a switchover between the fundamental and the first
overtone, indicated in Fig.~3 by the X linking columns 1 and 2.  There
is of course nothing special about the fundamental and first overtone, and we
see similar avoided crossings in Fig.~4 and corresponding switchovers in the
eigenvectors in Fig.~3 that are also indicated by crosses.

The lines with a positive slope in Fig.~2a correspond to modes with predominant
amplitudes in the outer well, i.e. with open tube characteristics, whereas the
those with negative slope have predominant amplitudes in the inner well,
i.e. are essentially closed tube modes.  It is thus the overlap of the open and
closed tube spectra that makes up the pattern of Fig.~2a.

The shapes of the fundamental eigenvectors in Fig.~3 show that changing $\zo$
is essentially equivalent to changing the length of the tube, and explain why
the periods increase/decrease linearly with $\zo$.

The periods $P_n$ change with $\zo$, but the ratios $P_{n0}$ are almost
everywhere independent of $\zo$ and thus straight lines.  The reason for this
near constancy is again that changing $\zo$ changes the equivalent length of
the tube.  We show the period ratios in Fig.~2b so that later they can be
compared to those of the Cepheid model sequence.

Why do the modes alternate between inner and surface modes?  Consider again
Fig.~3, column~1 starting at the bottom and going up.  Near the 
open end the eigenvector is
forced to bend down, at a constant rate because the curvature of $y$ is given
by $-k^2 y$.  Therefore the slope of the eigenvector in the outer region gets
increasingly steep as the barrier location moves inward.  Since there must be a
finite, fixed change in slope at $\zo$ (Eq.~12b) the slope of the eigenvector
of the inner region is increasingly flattened, thus suppressing the inner
amplitude (which has to go to zero at $\zo$=0).  The overtone modes follow the
same scenario.

We now examine more closely the case with parameter values $\zo=0.96$ and
$\Vo=10$ that approximately mimics the Cepheid situation.  Eq.~(13) is easily
solved and the the lowest 14 modes eigenvectors $y_n$ are displayed in Fig.~4
\vs acoustic depth $z$.  Referring back we see that this puts us on the extreme
left of Fig.~2.  We note in particular that the strange (surface) mode is
expected to appear only for large mode number.  Indeed, Fig.~4 shows that here
it is mode O11 that is strange\th (and mode O12 that is nearly matched).  Note
that for mode O12 a node occurs very close to $\zo$ which then causes no jump
in the slope (Eq.~13).  Modes higher than O12 therefore have at least one node
in $\zo$$<z$$\leq$ 1.

One can again look at the behavior of the modes in Fig.~4 in terms of matching
of the eigenvectors through the potential barrier.  Here now the curvature
increases with modal index (wave number $k$).  For the lowest modes the
curvatures and bending down of the eigenvectors are small and the modes are
inner (regular) modes.  As mode O9 is approached the matching has already
started to seriously suppress the inner amplitude which reaches a minimum
for O11 which is the surface (strange) mode.  Note that both O10 and O12 have
approximately the same inner and outer amplitudes and are hybrid modes.  For
mode O12 furthermore the node occurs very close to the barrier location $\zo$
and O12 is thus an almost perfectly matched mode.  The subsequent modes become
again regular modes, and through the same mechanism a new surface mode appears.
This whole scenario repeats itself with a relative recurrence period of
$\approx 1/(1-\zo)$ in the modal index (cf. Eq.~14).  

As a byproduct of the previous paragraph we note that the minimal amplitude and
the near match occur very close to each other, and that both phenomena are the
consequence of a near resonance between the inner and outer potential wells.

The connection with the usual stellar modes will become more apparent when we
consider the behavior of the $\delta r_n$ rather than the $\Psi_n$ in the next
section.

Thus in the Cepheid case the radial vibrational modes that are
confined to the surface regions have strange properties.  
Under cross-over conditions both modes are an admixture of normal and strange
modes.  This
is of course not all that different from what happens for nonradial Main
Sequence stars where different mode types, $g$ and $p$ modes, are involved in
the level crossings (\eg Unno \etal 1989).

\titlea{Details of a Cepheid Model}

 \begfig0cm
 \centerline{\psfig{figure=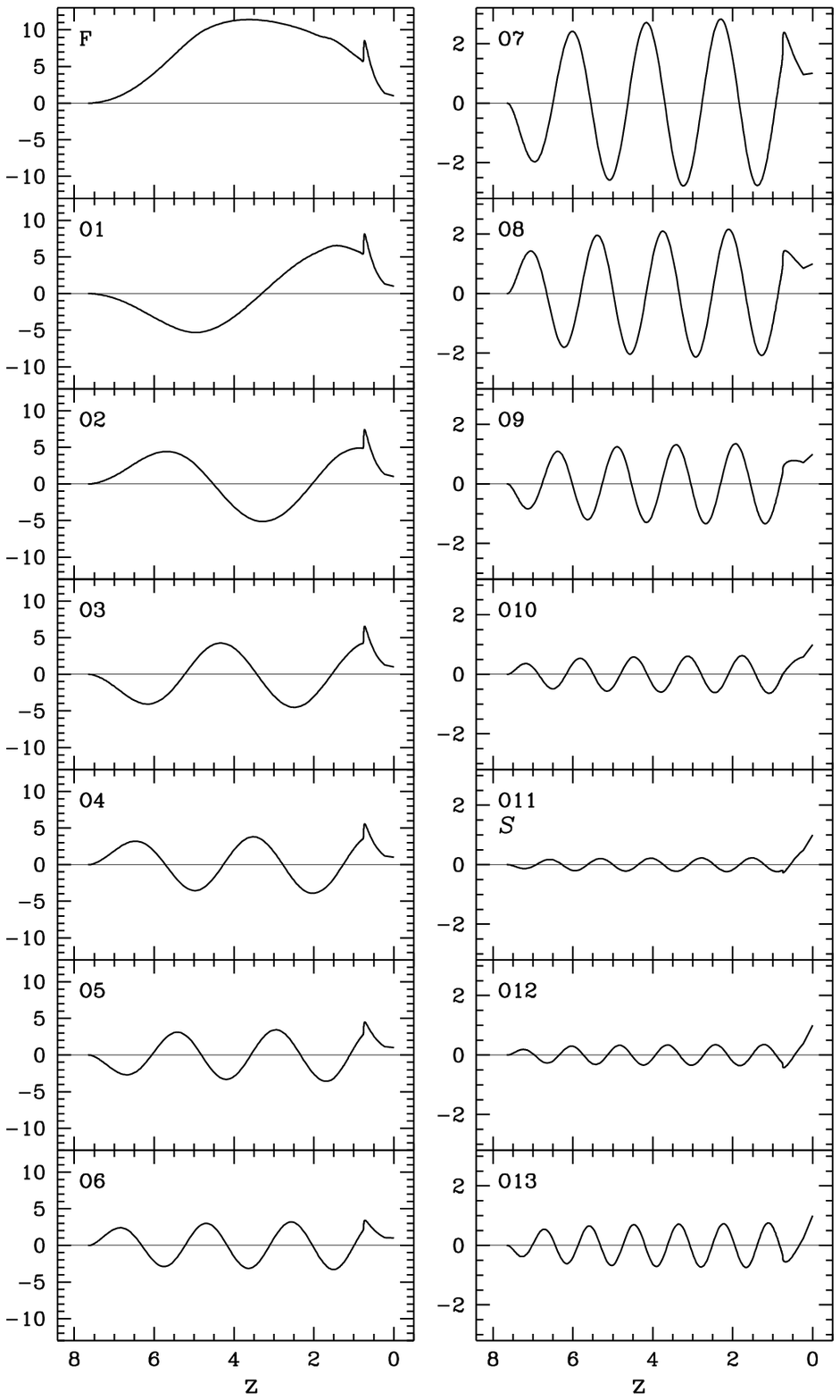,width=9.cm}}
 \vskip -10pt
 \figure {5}{Adiabatic eigenvectors of the scaled displacement $\Psi_n$ 
\vs acoustic depth
for the Cepheid Model; the strange mode is labelled by ${\cal S}$.
 }
 \endfig

 \begfig0cm
 \centerline{\psfig{figure=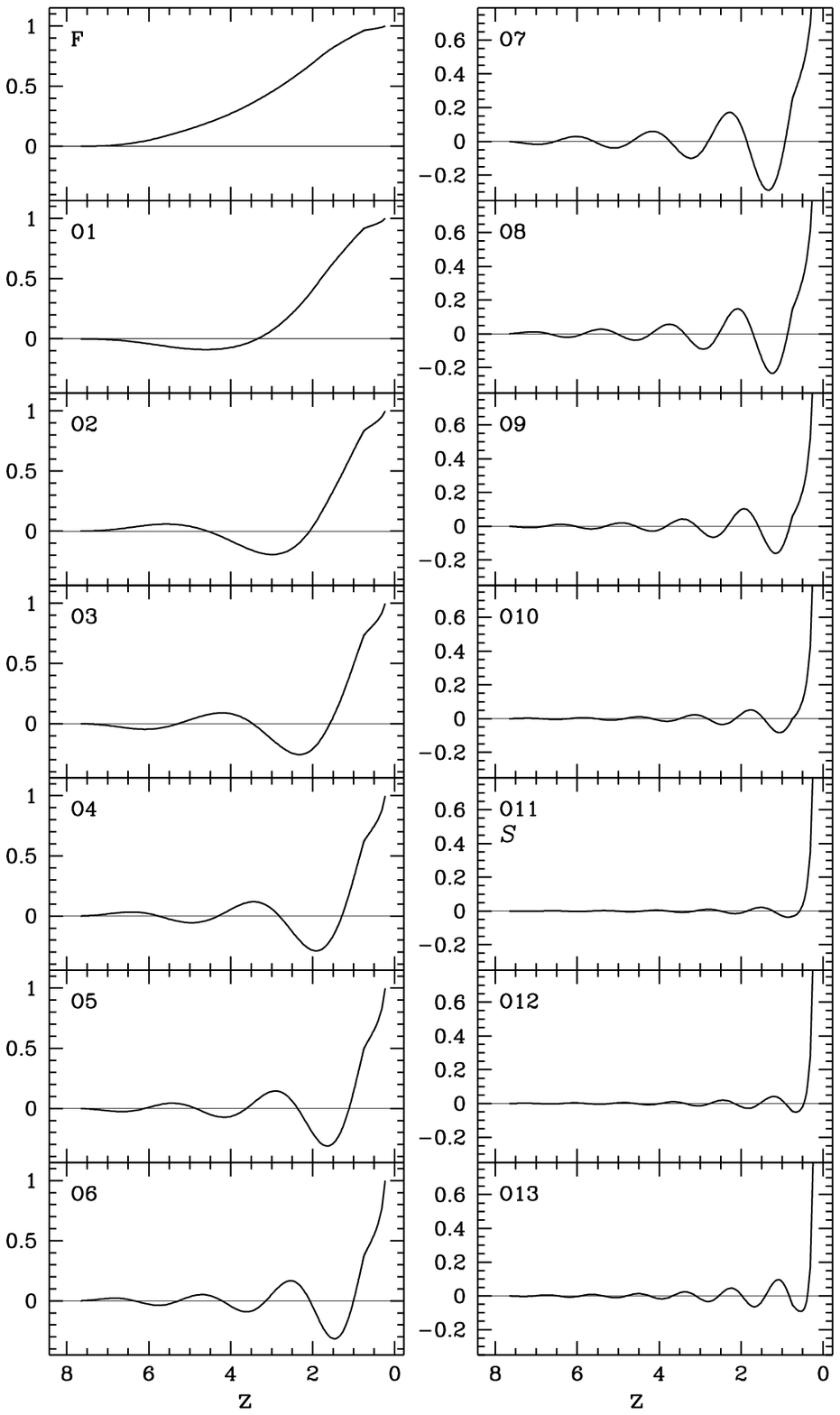,width=9.cm}}
 \vskip -10pt
 \figure {6}{Adiabatic eigenvectors $\delta r_n/r$ \vs acoustic depth
for the Cepheid Model; note the different scales on the left and on the right
column.  The strange mode is labelled by ${\cal S}$.
 }
 \endfig

Before discussing the strange modes in {\sl sequences} of nonadiabatic 
Cepheid models where they were first noticed (Buchler
1997) we first return to the Cepheid model of \S2 in order to show how
the toy problem of \S3 serves as a perfect paradigm for explaining the nature
of the strange mode.

We have to be a little bit careful with the outer boundary condition because of
the almost isothermal nature of the outer region.  In order to avoid problems
we define the surface as the point where the density is small, but nonzero,
or more conveniently where the pressure is equal to some small
value $p_*$, say two times the radiation pressure ($aT^4/3$).  We impose
this $p_*$ 
as a {\sl constant} outer pressure in the subsequent dynamics.  (The exact
value of this pressure is not critical).

\titleb{The Adiabatic Cepheid Model}

A great deal can be learned by examining the linear pulsations in the adiabatic
limit, \ie we ignore the entropy variations during the pulsation.

The adiabatic displacement 
eigenfunctions $\Psi_n$ for the first fourteen pulsational modes
are shown in Fig.~5, starting with the fundamental mode (F).  With increasing
$k$ a node moves outward toward $\zo$ and, as overtone mode O11 is approached,
the node is about to reach $\zo$.  The similarity with the analytical toy-model
of the previous section is striking.  As in the toy problem, the inner
amplitude decreases to a minimum, -- the property that has been used to
identify a strange mode.  There is no doubt that in an actual Cepheid model the
occurrence of the strange mode is also the result of the confinement of the
mode to the outer
cavity of the potential well.

There are some small differences with the toy problem, for example in the
detailed behavior both near $z=1$ and through the potential barrier.  This is
not astonishing considering that in the toy model we ignored the the potential
dip near $z=\zo$ as well as the broad, but relatively low potential barrier
near $z=1$ (The effect of the latter is seen in the exponential behavior of the
low modes near the surface.

It is interesting of course to compare the scaled eigenfunctions $\Psi_n$ to
the radial eigenfunctions $\delta r_n$ which are shown in Fig.~6.  The scale
factor $(r\sqrt{\rho c_s})^{-1}$ in Eq.~6 greatly magnifies the surface region
and the strange mode now appears even more as a surface trapped mode.

While the radial displacement eigenfunction $\delta r_n$ has perhaps a more
physical meaning, the $\Psi_n$ representation provides an optimal scaling of
the problem, and is perhaps more basic from a mathematical point of view.  It
shows in particular that even for a strange mode the inner regions play an
important role even though in $\delta r_n$ all modes appear to be 
surface modes.

We conclude that a {\sl strange adiabatic} mode has its origin in the potential
barrier that is created by the ionization front 
of hydrogen in the outer envelope.
It first appears when, with increasing mode number, an actual node is forced to
occur near the surface (outside the HPIR).  This then pushes down the inner
amplitude of the mode to a very low value.  The new, 
interesting point is that the
strange mode already appears in the adiabatic limit.

 \begfig0cm
 \centerline{\psfig{figure=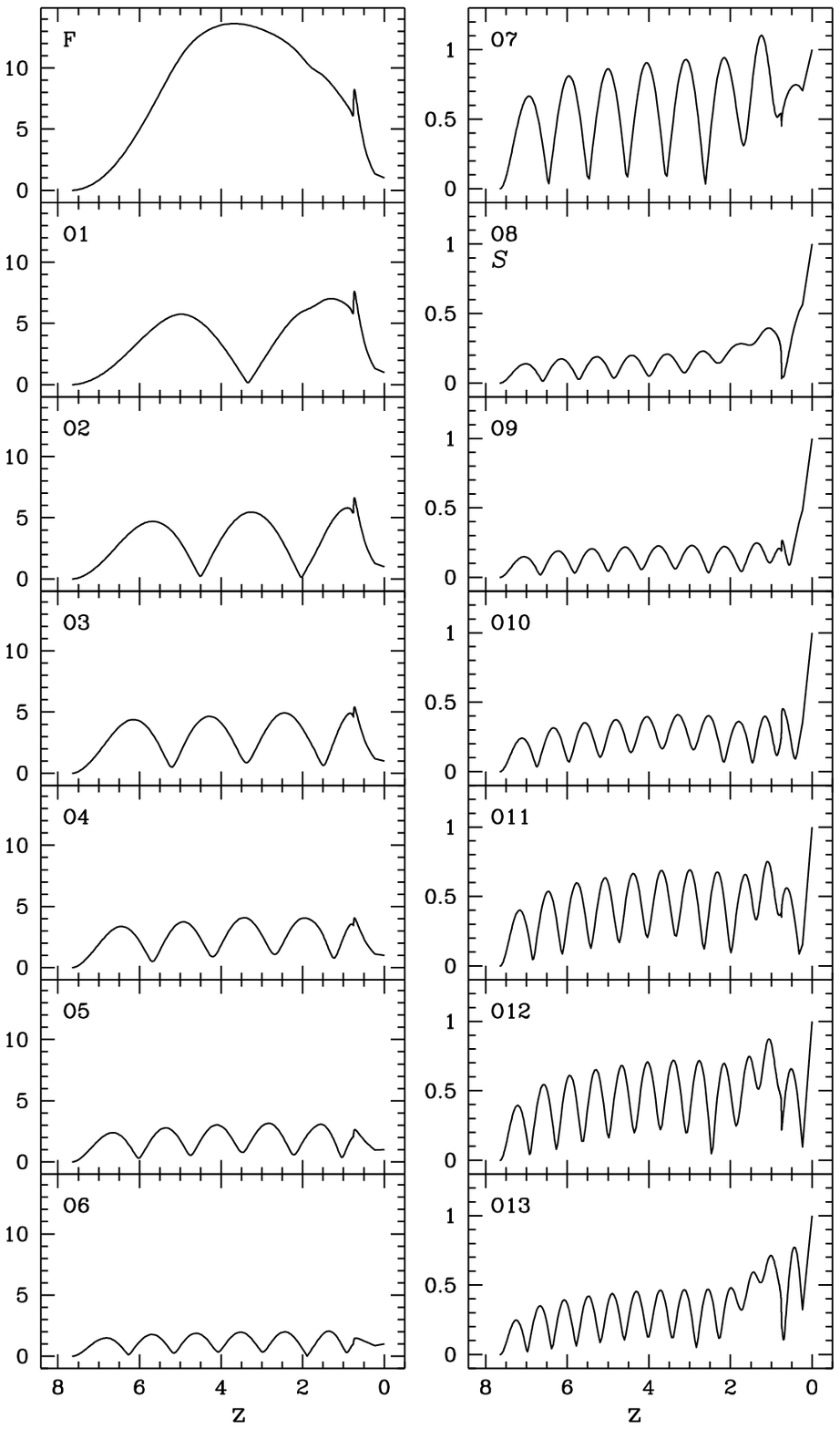,width=9.cm}}
 \vskip -10pt
 \figure {7}{Nonadiabatic eigenvectors $|\Psi_n|$ \vs acoustic depth for
the Cepheid Model; the strange mode is labelled by ${\cal S}$; note the
different scales on the right and left.
 }
 \endfig

 \begfig0cm
 \centerline{\psfig{figure=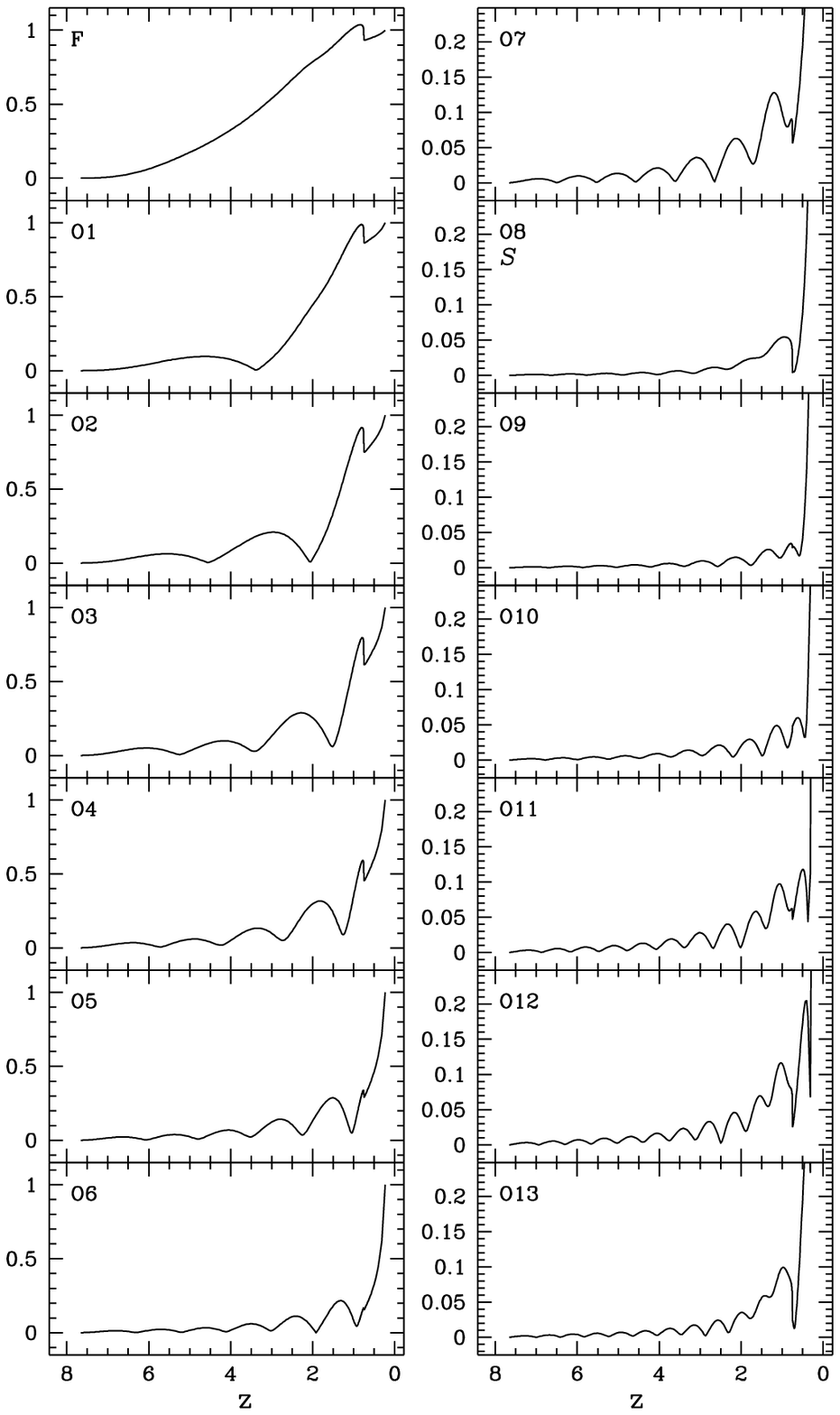,width=9.cm}}
 \vskip -10pt
 \figure {8}{Nonadiabatic eigenvectors of the scaled displacement
$|\delta r_n|/r$ \vs acoustic depth
for the Cepheid Model; the strange mode labelled by ${\cal S}$;  note the
different scales on the right and left.
 }
 \endfig

\titleb{The Nonadiabatic Model}

We now turn to the correct, {\sl nonadiabatic} linearization of the Cepheid
model.  In Fig.~7 we display the moduli of the {\sl nonadiabatic} 
displacement eigenvectors
$\Psi_n$, related to the radial eigenvectors through the scaling of Eq.~(6),
and normalized to 1 at the surface.  Juxtaposed we show in Fig.~8 the
corresponding radial displacement eigenvectors $\delta r_n/r$.  The
nonadiabaticity is weak for Cepheids ($\kappa_n \ll \omega_n$ for the low
modes) and therefore the properties of the nonadiabatic eigenvectors to a large
extent mimic those of the adiabatic problem.  \th\th (We recall that in our
notation the eigenvalues for an assumed ${\rm exp}(\sigma_n t)$ dependence are
denoted by $\sigma_n=i\omega_n +\kappa_n$.)

\titleb{Transition From Adiabatic to Nonadiabatic Model.}

The nonadiabatic spectrum is noticeably different from the adiabatic one as
Table 1 shows, in particular it is mode O8 which is now the strange mode
(denoted by the symbol ${\cal S}$).  It is naturally of interest to explore how
the transition from adiabatic to nonadiabatic occurs.  To that effect we show
in Fig.~9 the period ratios as a function of the logarithm of the
nonadiabaticity parameter $\epsilon$ \th\th (We use the same definition as
Zalewski 1992; \viz in the dynamical equation $\epsilon$ multiplies the entropy
contribution to the pressure variation in the mechanical equation, $\epsilon$
=0 thus being the adiabatic case and $\epsilon$=1 being the full nonadiabatic
one).

Note that it is indeed for the strange modes that the period ratio changes most
rapidly in Fig.~9.  The reason for this extreme sensitivity is also clear.  The
resonance condition is very sensitive to the outer region.  But there the
Cepheid envelope is highly nonadiabatic (\eg Cox 1980) even though, overall,
the Cepheid pulsation is weakly nonadiabatic, as already pointed out.  It is
therefore expected that even a small nonadiabaticity can have the dramatic
effect of gradually 
shifting the strange mode from O11 down to O8 as Fig.~9 indicates.

 \begtabfull
 \tabcap{1} {Comparison of exact and adiabatic periods\hfill\break From left to
right: mode, period, growth-rate, moment of inertia, adiabatic period; the
strange mode is indicated with {\cal S}.}
 \halign{
 \hf\quad\enskip\th #\enskip\quad\hf&
 \hf\enskip #\enskip&
 \hf\enskip #\enskip&
 \hf\enskip #\enskip\th\th\th\th\hf&
 \hf\enskip #\enskip&
 \hf\enskip #\enskip\cr
 \noalign{\smallskip\hrule\medskip}
  mode&  $P_n$[d]  &     $\eta_n$  & $I_n$ &  
 $P^{ad}_n$[d] \cr
 \noalign{\smallskip\hrule\smallskip}
  F &   9.089 &   0.018 & 2.10e+30   & 9.145  \cr
 O1 &   6.278 &   0.054 & 4.71e+29   & 6.324  \cr
 O2 &   4.525 &  --0.121 & 4.18e+29   & 4.507  \cr
 O3 &   3.554 &  --0.291 & 4.39e+29   & 3.517  \cr
 O4 &   2.937 &  --0.371 & 3.29e+29   & 2.885  \cr
 O5 &   2.494 &  --0.313 & 2.06e+29   & 2.437  \cr
 O6 &   2.164 &  --0.207 & 9.42e+28   & 2.109  \cr
 O7 &   1.907 &  --0.100 & 2.21e+28   & 1.853  \cr
 O8 &   $^{\scriptscriptstyle S}$1.729 &   0.209 & 1.60e+27   & 1.650  \cr
 O9 &   1.649 &  --0.261 & 1.88e+27   & 1.489  \cr
 O10&   1.534 &  --0.319 & 5.38e+27   & 1.362  \cr
 O11&   1.400 &  --0.232 & 1.55e+28  &$^{\scriptscriptstyle S}$1.281 \cr
 O12&   1.285 &  --0.215 & 1.76e+28   & 1.219  \cr
 O13&   1.193 &  --0.177 & 8.70e+27   & 1.139  \cr
 O14&   1.135 &  --0.181 & 4.88e+27   & 1.063  \cr
 \noalign{\smallskip\hrule}
 }  
 \endtab

\titlea{Cepheid Sequence}

 We now turn to a Cepheid model sequence with $M$=5\Mo, $L$=3000\Lo, $X$=0.7,
$Z$=0.02, in which the variable parameter is the fundamental period $P_0$, or
equivalently the \Teff\ ranging here from 8600 K to 4600 K.  We have chosen
this somewhat overluminous sequence (on the Chiosi $M$--$L$ relation) simply
because the strange mode appear as a vibrational overtone with
lower modal number.

\titleb{Modal spectrum}

We start again by examining the adiabatic models, and show in Fig.~10 the
behavior of the {\sl adiabatic} period ratios $P_{n0}=P_n/P_0$ for the model
sequence.  The adiabatic modes are seen to display avoided level crossings.
Here we can prove that the levels cannot cross.  Indeed, because of the
Sturm-Liouville nature of the adiabatic eigenvalue problem (\eg Ledoux \&
Walraven 1958) the modes have distinct frequencies -- no degeneracy and no
crossings.

 \begfig0cm
 \centerline{\psfig{figure=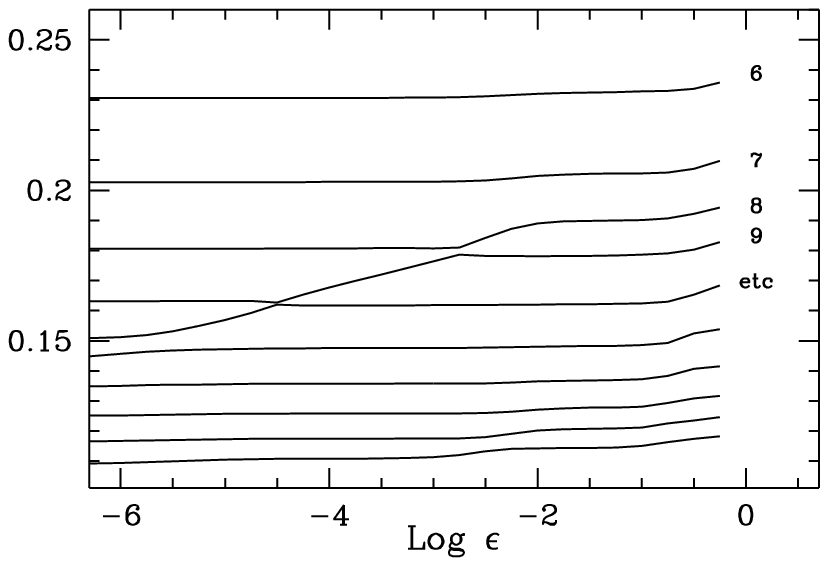,width=9.cm}} 
 \vskip -10pt
 \figure {9}{Period ratios $P_{n0}$ in the transition from adiabatic to
 nonadiabatic model as a function of Log$(\epsilon)$, where $\epsilon$ is
 the nonadiabaticity parameter}
\endfig

 \begfig0cm
 \centerline{\psfig{figure=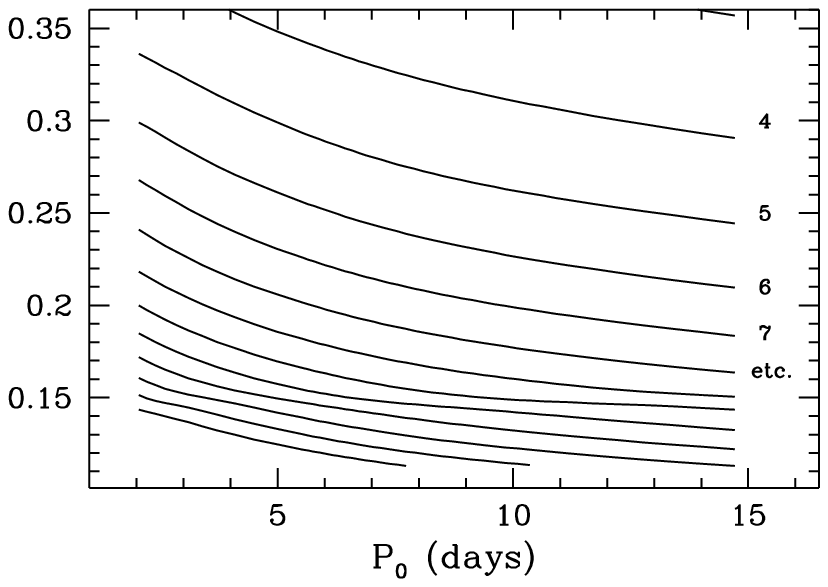,width=9cm}}
 \vskip -5pt
 \figure {10}{Cepheid model sequence; Period Ratios $P_{n0}$ of the {\sl
adiabatic} spectrum based on the first fifteen overtone modes of the spectrum
(k=1, 15); only ratios with $k>3$ are shown.
 }
 \endfig

Guided by the toy problem (Figure~3) we understand that the level crossings are
a consequence of the relative shift in the location of the potential barrier,
and thus in the resonance condition.  Indeed with increasing period (or
decreasing \Teff ) the HPIR moves inward, causing the strange mode to move to
lower modal number.  As we go from high to low \Teff\ (low to high $P_0$) the
HPIR moves inward, and concomitantly the outer potential well increases in
width relative to the inner one.  In the toy problem this corresponds to
decreasing $\zo$.  We therefore anticipate a behavior comparable to that of the
toy problem displayed in Fig.~3.

Turning now to the full {\sl nonadiabatic} models we plot in Fig.~11, on top,
the period ratios which also show avoided crossings However, they are shifted
with respect to the nonadiabatic picture and are
more dramatic.  The mode labelled
15 on the left successively trades places with modes 14, 13, 12, etc and ends
up as mode 7 on the right side of the figure.

 \begfig0cm
 \centerline{\psfig{figure=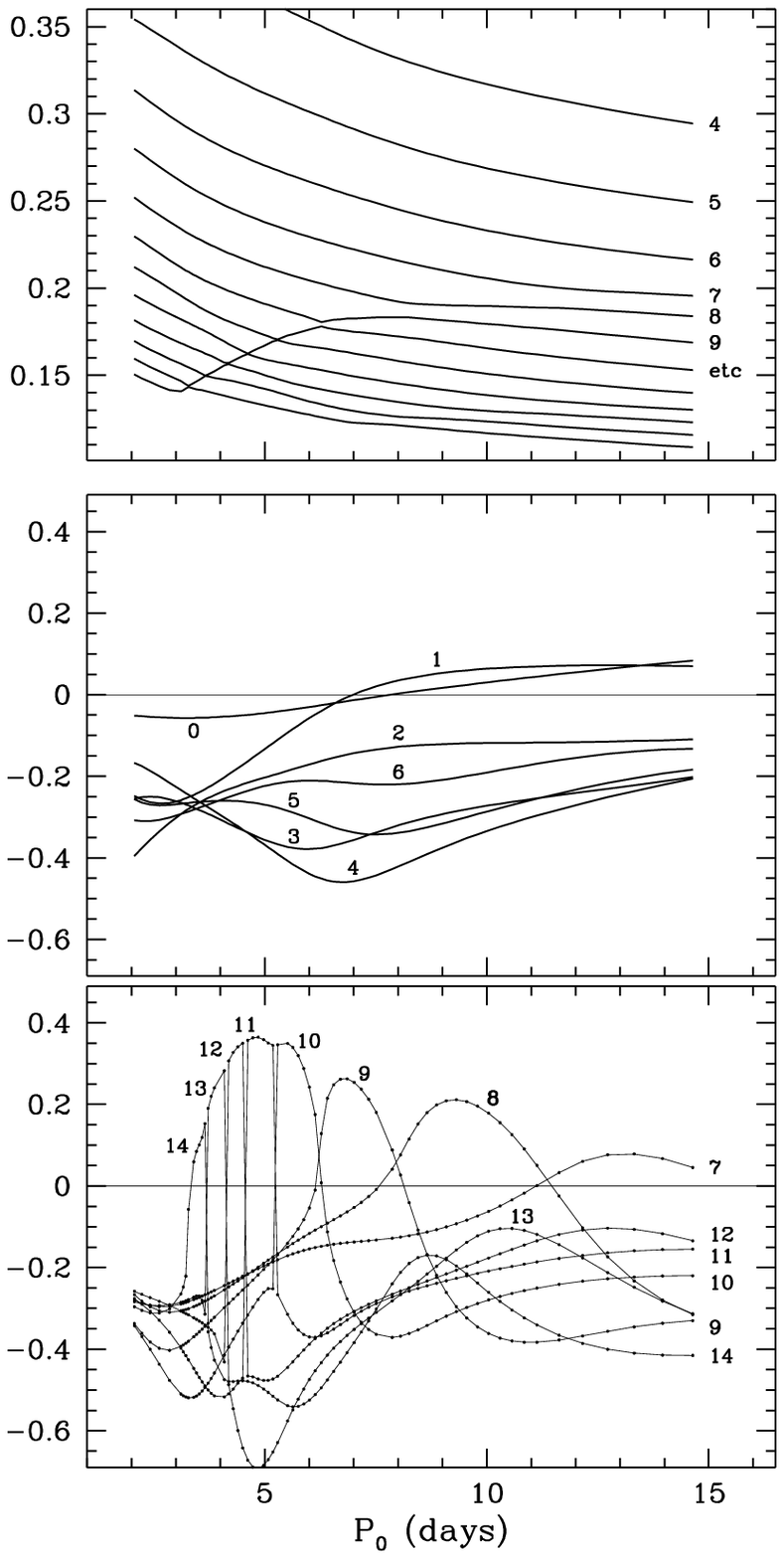,width=9cm}}
 \vskip -10pt
 \figure {11}{Nonadiabatic Cepheid model sequence, {\sl top:} Period ratios,
$P_{n0}$,k=1,14; {\sl middle:} relative growth-rates, $k$=0,6; {\sl bottom:}
relative growth-rates, $k$ =7, 12; individual models are indicated by points in
the bottom graph.
 } 
 \endfig

The middle and bottom graphs in Fig.~11 display the relative growth-rates
$\eta_n = 2\kappa_n P_n$, for modes $k$=0 through 6, and $k$=7 through
14, respectively.  Thus the first overtone in this sequence is unstable for
\Teff $<$ 6000\th K ($P_1$=4\dotd 7) and the fundamental for \Teff $<$ 5800\th
K ($P_0=$ 7\dotd 7).

 \begfig0cm
 \centerline{\psfig{figure=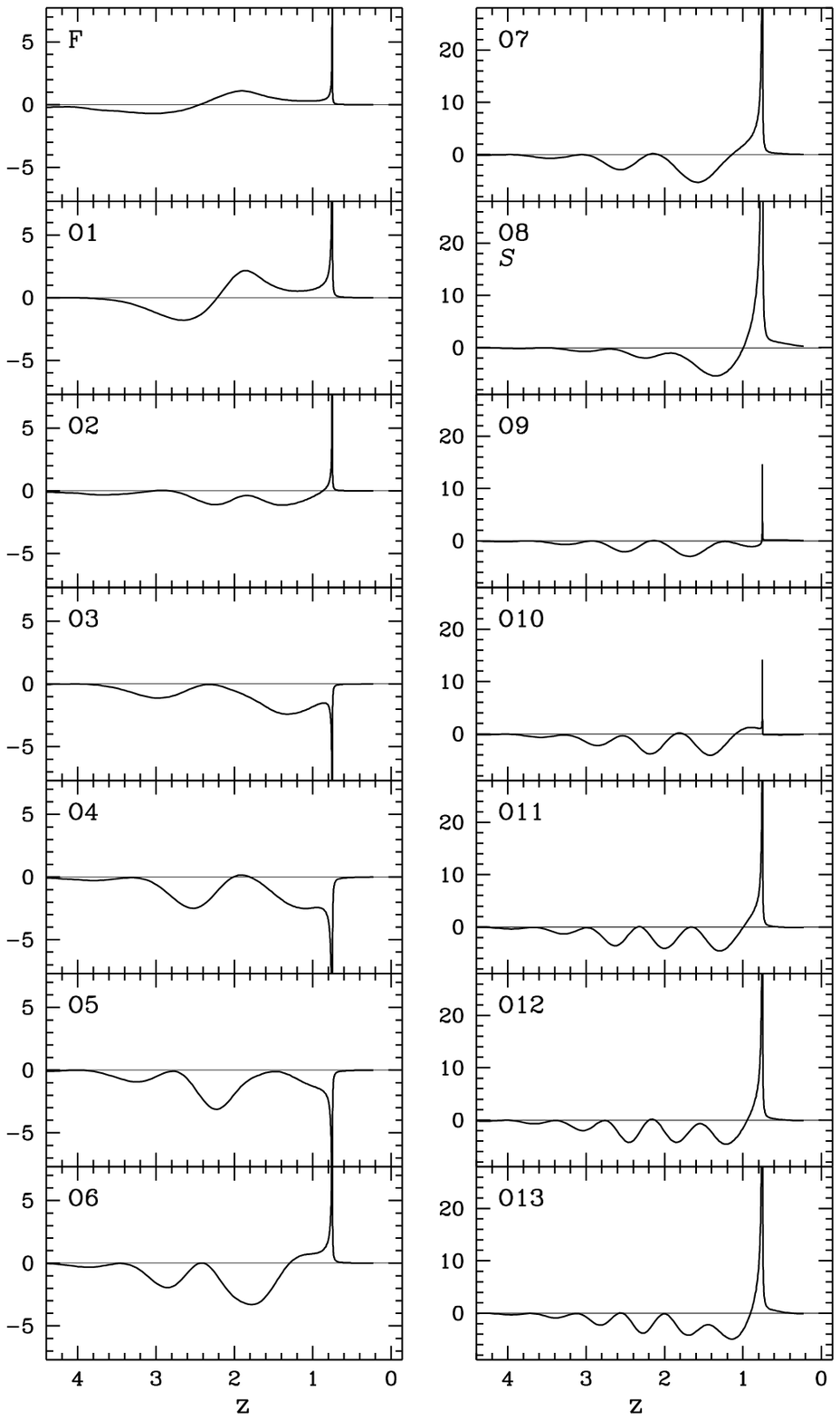,width=9.cm}}
 \vskip -10pt
 \figure {12}{Work integrands \vs acoustic depth for
Cepheid Model; the strange mode is labelled by ${\cal S}$.
 }
 \endfig

In the bottom graph that shows the higher overtones we witness some interesting
behavior.  The lines connect the points ordered by decreasing $P_{n0}$ which is
different from the way the eye connects them.  There is an unstable 
modal feature
throughout the whole period range, but it is not always the same mode.
Furthermore whenever a level crossing occurs in the top graph, there is a
corresponding switch-over in the growth-rates of the two corresponding modes.
The switchover is relatively slow for modes 7 to 8 and 8 to 9, more rapid for
modes 9 to 10 and 10 to 11 and very rapid for the higher modes.  (The reason is
simply that the position of the HPIR moves more rapidly with $P_0$ for low
$P_0$.)  In both the period ratios and in the growth-rates it is as if some
egregious modal property hops  from mode 14 to mode 7.

A more careful inspection of Fig.~11 uncovers among the higher modes the
appearance of a second strange mode in the period ratios.  Some secondary
structure is even more clearly visible in the growth-rates.  There are local
maxima for modes 13 and 14 at 10\dotd 5 and 8\dotd 5 days, respectively, which
are due the appearance of higher order strange modes -- by which it is meant
that a second node crosses the HPIR.

In a recent paper Glasner \& Buchler (1993) already noted that for RR Lyrae and
Cepheid models the stability of the vibrational modes does not increase
monotonically with mode number.  Instead, there is a strong excursion back
toward instability around the eighth or ninth overtone.  At the time they had
no physical explanation for the nonmonotone behavior, but the present results
show that this is related to the recurrent appearance with increasing modal
number of strange modes.

\titleb{Work-integrands}

Fig.~11 shows that the strange mode has a very different growth-rate from its
neighbors and it is interesting to inquire why.  In Fig.~12 we therefore
examine the work integrands, \ie the work $\langle p dv\rangle$
done per cycle, or the integrand of the expression for the relative
growth-rate
 $$\eqalign{\eta_n = 
  2{\kappa_n\over \omega_n} 
    &={2\pi \over \omega_n^2 I_n}{\cal }Im \int \delta p_n\delta v_n^* \th
4\pi r^2\rho c_s \th dz \cr
    &={2\pi \over \omega_n^2 I_n} \int \Delta p\Delta v 
          \th {\rm sin} (\phi_p-\phi_v) \th 4\pi r^2\rho c_s \th dz \cr
 }\eqno(15)
 $$
 \ni where the $\phi_p$ is the phase of $\delta p_n$ and $\Delta p_n$ its
modulus, and similarly for the specific volume eigenvector $\delta v$.  
The quantity $\eta_n$ represents the energy growth of a mode over one
period -- its inverse is equal to the quality factor $Q_n$ commonly 
associated with resonant electronic devices.
The quantity $I_n$ is the moment of inertia of the mode:
 $$I_n = \int \vert\delta r_n\vert^2 \th dm
 \eqno(16)
 $$
 \ni (\eg Glasner \& Buchler 1993). A positive work-integrand means that
locally internal energy is converted into kinetic (pulsational) energy.

In Fig.~12 we exhibit the work-integrands for the lowest 14 modes as a function
of $z$ \th (The areas under the curves are the relative growth-rates).  It is
apparent in Fig.~12 that the egregious growth-rate of the strange mode is the
result of two effects.  First, the inner pulsation amplitude that we have shown
to be characteristic of the strange mode causes both $\Delta p$ and
$\Delta\rho$ and hence the integral (eq.~15) to be much smaller for the strange
mode.  Thus the surface region plays a relatively larger role in the
growth-rate.  Second, the smallness of the integral is compensated for by the
considerable amplification that comes about from the small moment of inertia
$I_n$ in the denominator of Eq.~15.  The resulting strange growth-rate is
therefore largely the result of the contribution from the outer region.  In
this example, the outer region happens to be driving, this is the source of the
positive growth rate of the strange mode.

Since for the strange modes almost all of the driving comes from the HPIR the
stability of these modes is therefore more sensitive to the physical and
numerical resolution of the outer layers.  However, as we have seen, 
their existence
and nature are not sensitive at all.

More light can be shed on the behavior of the growth-rates by examining a
non-adiabatic extension of our toy model, which we consider next.

\titlea{Nonadiabatic Toy Potential Barrier Problem}

The nonadiabatic problem is of course much more complicated, but it 
is possible to extend the potential barrier problem of \S2 to mimic
nonadiabatic pulsations by generalizing Eq.~9:
 $$-\ty'' + {\cal V}\th \tk\delta (z-\zo) \ty +iU \th \ty = \tomega^2 \ty
 \eqno(17)$$ 
 where $\tk$ is the (complex) wave-vector $\tk=\sqrt{(\tomega^2-iU)}$,
and $\tilde \omega$ is the complex frequency.  The
potential $U$ mimics an overall dissipation.  In the adiabatic case, ${\cal V}$
was a real constant of value $\Vo$.  Here we extend the potential to include an
imaginary component ${\cal V}=\Vo + i\Delta \Vo$.  The choice of $\Delta \Vo$
represents a crude solution of the nonadiabatic pulsation equations:
 $$
\tomega^2 \delta r = {\rm G_1} \delta r + {\rm G_2} T \delta s \; ,
 \eqno(18a)$$
 $$
i\tomega T\delta s = {\rm K_1} \delta r + {\rm K_2} T \delta s \; ,
 \eqno(18b)$$
 in which we assume that $({\rm G_2}( i\tomega - {\rm K_2})^{-1}{\rm K_1})\th
\delta r$ can be replaced with a complex term ${\cal B}\delta r$, proportional
to $\delta r$ and thus to $y$.
It has been recognized that the shape of the entropy fluctuations is
independent of the mode (Pesnell \& Buchler 1986) and is well represented by a
sharp peak in the ionization region.  As a result we superimpose the complex
$\Delta \Vo$ onto the adiabatic delta function potential.

 \begfig0cm
 \centerline{\psfig{figure=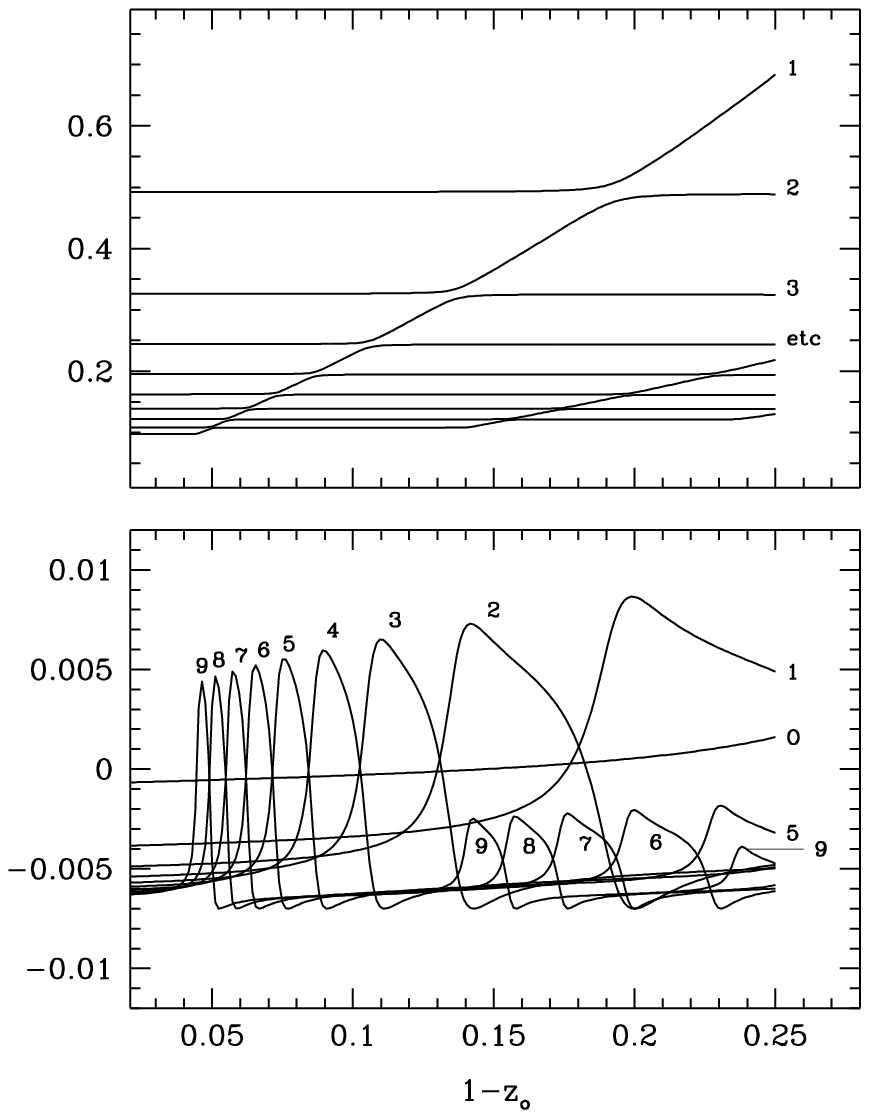,width=9.cm}} 
 \vskip -10pt
 \figure {13}{Period ratios (top) and relative growth-rates (bottom) as 
  a function of outer well width $1$-$\zo$ for the lowest 9 modes of the
complex barrier problem with ${\cal V}=10-2i$, $U$= 0.02.
 }
 \endfig

 The inner boundary condition is still $\ty(0) = 0$, but the outer one can now
be more general (\eg Morse \& Feshbach, 1953, Eq.~11.1.21~ff), \viz $\delta p =
R \dot u + X u$ which translates here into $\ty'(1) = {\cal Z} y(1)$, where
${\cal Z}$ is a complex
surface impedance.  ${\cal Z}=0$ corresponds to a free surface
as in the adiabatic case (perfect reflection).  ${\rm Im} {\cal Z}$ is
the inertia presented to the stellar surface by an overlying atmosphere, and
${\rm Re} {\cal Z}$ corresponds to acoustic radiation losses.

For highly nonadiabatic stars the adiabatic potential
barrier that is contained in ${\rm G}_1$ becomes less important relative to the
seond term, and it is the
$\delta $ function from the peaking of $\delta s$ that dominates the problem.
We note in passing that Zalewski (1992) decided to include an inhomogeneous
term $A\delta(z-\zo)$ instead of a potential, and took ${\cal Z}$ to be a real
constant $v_0$ (whose value he did not specify in the paper).  His simple
model does not work for Cepheids although it matches some, but not all the
properties of the spectrum of the highly nonadiabatic PAGB models.

 With the appropriate matching conditions $y(\zo^+)=y(\zo^-)$ and
$\ty'(\zo^+) = \ty'(\zo^-) + \tk \th {\cal V}\th \ty(\zo)$,
 one obtains the eigenvalue equation for the complex wave-vector $\tk$
 $$\eqalign{
  \th & \cos \tk -\ngth {\cal Z} \sin \tk \cr
  & = -\Vo \th \sin \tk \zo
 (\cos \tk (1-\zo) - \ngth{\cal Z} \th \sin  \tk (1-\zo)) \cr
 }\eqno(19)$$
The solution is
given by
 $$\vcenter{\openup1\jot
   \halign{ $\hfil#$  &  ${}#\hfil$  &  \th\th\quad \quad $\hfil#$ 
                      &   ${}#\hfil$   \cr
 \ty(z) &= a\th  \sin \tk z                        & z \leq \zo\cr
      &= \cos  \tk (1-z) - {\cal Z} \sin  \tk (1-z) & \zo \leq z\cr}
 }\eqno(20)$$
 where the inner amplitude $a(\tk)=y'(0)/\tk y(1)$ is now
 $$a(\tk)= (\cos \tk (1-\zo) -{\cal Z} \sin \tk (1-\zo))/ \sin \tk\zo
 \eqno(21)$$

Again, we examine a sequence of toy models, but this time for a relatively
narrow range of barrier locations, $\zo$.  We have chosen the parameters
${\cal V} = 10 - 2i$ and $U=0.02$.  We set ${\cal Z}=0$ consistent with $p_*
\approx 0$, as in the hydrocode.  The period ratios for this sequence are shown
in Fig.~13 (top) and it is easy to follow the strange property from overtone 9
at the bottom left corner to overtone 1 at the top right.  In Fig.~13 (bottom)
we show the corresponding growth-rates of the lowest modes, as obtained with
Eq.~19.

Considering the crudeness of the toy model's $\delta$ function potential, the
properties of the growth-rates mimic those of the Cepheid sequence remarkably
well.  Here, as in Fig.~11, we see the increasingly rapid switchover of the
unstable excursions and even, at bottom right, the weaker excursions of the
second and third order strange modes.  In this toy model, the effect of second
and third order strange modes are also visible in the period ratios 
(Fig.~13 top)
as it was
in the Cepheid sequence (Fig.~11).

The behavior of our eigenvalues can actually be understood analytically via
perturbation theory when the nonadiabaticity is small.  The correction
$\Delta \omega^2$ 
(where $\tomega^2 = \omega^2 + \Delta\omega^2$ to first order) to
the eigenvalue due to the complex potential $U$ and the small imaginary
contribution to ${\cal V}$ is given in terms of unperturbed (adiabatic)
quantities
 $$\eqalign{
 \Delta\omega^2 &= i U + i\Delta \Vo k y^2(\zo)/{\cal I} \cr
             &= i U + i\Delta \Vo a(k)^2 \sin^2 k \zo/{\cal I} \cr
 }\eqno(22)$$
 \ni where ${\cal I} = \int \vert y\vert^2 dz$.  The correction is seen to be
at a {\sl maximum} for the {\sl strange} mode for which 
${\cal I}$ is at a minimum.
In contrast, it is at a {\sl minimum} for the neighboring {\sl matched} mode
for which $y(\zo )\approx 0$.  Ths explains the rapid variation of the 
growth-rates in the vicinity of the strange modes 
Fig.~13 confirms both of these predictions.  The
potential $U$ by construction provides an overall shift of $\Im\th\omega$ to
negative values.

\titlea{The Cepheid Horn}

The reduction of the adiabatic pulsation problem to a Schr\"odinger like
equation is very well known in the study of wind instruments (Morse \& Ingard
1968, Benade 1977).  But it is common to focus on pressure perturbations,
though it is straightforward to show the equivalence of the following two
descriptions of linear sound waves in a horn, the first for the velocity
perturbation $u$ (or equivalently for $\delta r$), 
and the second, well known Bernoulli-Webster equation, for
the pressure $\delta p$ 
 $$\eqalignno{
 {1\over \rho}
 {\partial\over\partial x}\left( {\rho c_s^2\over S}
 {\partial \over\partial x} (Su) \right) &= \omega^2 \th u \th ,
 &(23)\cr 
 {c_s^2 \rho\over S}
 {\partial\over\partial x}\left( {S\over\rho}
 {\partial  \over\partial x} \delta p \right) 
        &= \omega^2  \delta p\th ,
 &(24)\cr
 }$$
 In the case of an acoustic horn it is not usually necessary to consider
gradients in the sound speed $c_s$ and density $\rho$, and it is {\sl the shape
of the spatially varying cross-sectional area} $S(z)$ that determines the
potential $V(z)$.

 Eq.~(23) can be transformed into a Schr\"odinger equation of the form (Eq.~7)
in the same way as Eq.~(1), in which the potential is
 $$
 V_u(z) = \left(S/\rho\right)^{\smhalf}
 {d^2\over dz^2} \left(S/\rho\right)^{-\smhalf}  \; .
 \eqno(25)
 $$ 
 {\sl Mutatis mutandis}, the potential for Eq.~(12) is shown to be
 $$
 V_p(z) = \left(S/\rho\right)^{-\smhalf}
 {d^2\over dz^2} \left(S/\rho\right)^{\smhalf}  \; .
 \eqno(26)
 $$
 The potential $V_p(z)$ is the so called {\sl horn function} (Morse \& Ingard
1968, Benade 1977).  (We note in passing that a standard reference to the
Webster-Bernoulli equation, \viz Eq.~1 in Eisner (1967) incorrectly uses
Eq.~(24) instead of Eq.~(23) as the equation for the displacement.)

In the case of the horn it is the curvature associated with the flaring that
can give rise to potential barriers.  Reflection at these barriers (often
phrased in terms of acoustic impedance mismatch) produces internally trapped
modes, which are the discrete tones of the horn or wind instrument in question.
In contrast, loudspeaker horns are flared exponentially so as to avoid
potential barriers.

Can we design a wind instrument with a flare function ${\cal R}(z)=\sqrt{S(z)}$
chosen so as to have the same acoustic spectrum as our variable star?  In the
Cepheid, as we have seen, the potential is produced by the variation of $\rho$,
$c_s$, the spherical geometry, and by gravity.  Can we work backwards from the
potential of Eq.~(8) to find the shape ${\cal R}(z)$ of horn that would
reproduce a Cepheid's tones?

\begfig0cm
\centerline{\psfig{figure=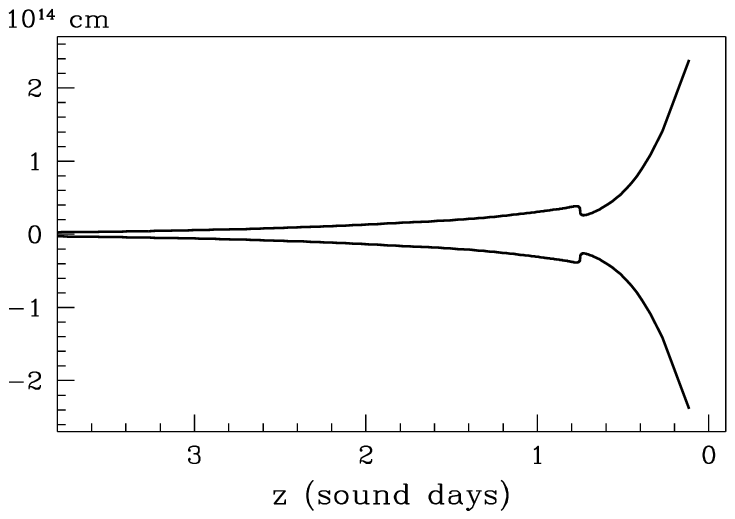,width=9.5cm}} 
\vskip -10pt
 \figure {14}{Cross-section of the horn corresponding to Cepheid
 pulsations, shown over the outer potion of the pulsating envelope
 (The entire model envelope extends to $z=8$); the kink at $z=0.75$ 
 corresponds to the potential barrier.}
\endfig

A closer inspection of the stellar problem shows immediately a big difference
with a horn, aside from the strong spatial gradients of density and sound
speed: The potential $V$ in Eq.~(8) contains 
a gravitational term that cannot be absorbed away into the term that has the
form of a horn function.
The presence of the
gravitational potential term thus 
makes the inversion from the potential to the
flare function impossible unless a Cowling approximation is made.

With the Cowling approximation we can construct an approximate equivalent
acoustic horn -- with the horn part being of primary importance everywhere
except in the innermost regions of the star.  In Fig.~14 
we show a slice through
the horn that would produce the same horn function as the Cepheid model (with
\Teff=5520 K).  The modes produced by this horn of course have somewhat
different periods from the original Cepheid because of the Cowling
approximation.  It is indeed well known that this approximation is not very
accurate for low radial modes, and the reason has already been mentioned: The
nodal structure is determined by the whole envelope, as the transformation to
$z$ has shown particularly clearly.  Nevertheless, it is amusing to see that in
the region where the horn function dominates the potential, a constriction in
the bell of the horn arises because of the HPIR.

Because the strange modes are trapped surface modes they should feel very
little the gravitational potential.  They can therefore well described with
the horn potential only.  Interestingly the radial displacement eigenvectors
have an appreciable amplitude only the outer, open part of the flare.

\titlea{Nonlinear Pulsations}

The strange modes are seen to be unstable to the left of the blue edge of the
fundamental mode and the first overtone -- in fact the strange mode can be the
only unstable vibrational mode there (Fig.~11).  It is of course interesting to
investigate 
what the pulsations of this star look like and why no observational evidence
has been reported for variability in this region of the HR diagram.

We have seen in Fig.~8 that the region outside the HPIR dominates the
radial eigenvectors and thus the pulsation.  {\sl A priori} we have no idea of
what the amplitude of pulsation might be.  We have therefore computed the
nonlinear (limit cycle) 
pulsations of several models from our sequence, that are located to
the left of the Cepheid instability strip (and are therefore not Cepheids {\sl
stricto sensu}).

For all these models the limit cycle 
pulsations have a
very small amplitude, typically 0.1 -- 1 km/s photospheric 
velocity and magnitude
variations of milli-mags.  If such 
'strange Cepheids' indeed exist they are
very hard to detect.  Aikawa \& Sreenivasan (1995) have found similarly small
amplitudes for yellow supergiant and intermediate mass stellar models pulsating
in a strange mode.

The effects of the strange mode might also be felt directly in the regime of
'normal' Cepheids.  Two simple scenari are possible here 
in theory: (a) the strange mode is
linearly unstable together with the fundamental (or with the first overtone);
if the nonlinear coupling coefficients happens to have suitable values, then 
beat pulsations can occur
(The conditions that must be satisfied are described in great detail in Buchler
\& Kov\'acs 1986); (b) the strange mode is resonant through a higher order
resonance, say 4:1 or 5:1, cf. Fig.~11, top).  If the strange mode is further
linearly unstable or just marginally stable and the nonlinear coupling
coefficients are right, then periodic, frequency locked pulsations can occur
(irregular amplitude modulations are also possible in principle, but less
likely).
Finally, the strange modes could have an indirect effect in that they could
destabilize the fundamental (or first overtone) limit cycles in a regime where
the first overtone (or fundamental) limit cycles are also unstable.  The result
would then be a usual beat Cepheid, i.e. constant amplitude pulsations in
both the fundamental {\sl and} first overtone (Buchler 1997).

For the strange modes to be able to have any of the described effects depends
on the strength of the nonlinear coupling coefficients.  Whether the small
overlap between the normal and the strange modes allows a strong enough
coupling is unknown at present.  Hydrodynamic computations are in progress to
explore these phenomena.

\titlea{Conclusions}

We have shown that the problem of adiabatic radial pulsations can be very
instructively recast in terms of a Schr\"odinger equation -- this without any
approximations.  The sharp variations in sound speed through the hydrogen
partial ionization region create a very high and very narrow potential barrier,
separating the pulsating envelope into two regions.  The appearance of strange
modes is a consequence of the possibility of mode-trapping 
in the surface regions.

While this Schr\"odinger equation formulation is stricty correct only in the
adiabatic limit, the same phenomenon of trapping persists when nonadiabatic
effects and acoustic radiative losses are taken into account.

It is clear from our study that the strange modes are thus of acoustic origin,
rather than arising as a coalescence of two thermal or secular modes.

The phenomenon of trapping has been illustrated with a simple analytical toy
problem that exhibits all the important features associated with the strange
modes, such as avoided level crossings.  This model can also be extended to
explain the properties of the nonadiabatic spectrum.  The resemblance with the
behavior of the periods and growth-rates of a Cepheid model sequence is
striking.

There is of course no reason why strange modes should not occur in the
nonradial spectrum of low $\ell$ $p$ modes, because these modes feel
essentially the same potential as the radial modes (although a Schr\"odinger
formulation is no longer possible without approximation because of the higher
order of the nonradial equation).  The level crossings are therefore distinct
from those that occur when $p$ and $g$ modes interact (Unno et al. 1989).

Finally, we have examined the nonlinear pulsations associated with the strange
modes and have concluded that they have very small amplitudes, especially in
luminosity.

\vskip 30pt

We wish to thank Art Cox for his comments on this paper. 
This work has been supported in part by NSF (grants AST92-18068, INT94-15868 
and AST95-28338).

\vskip 15pt

\begref{References}

\ref
 Aikawa, T. 1985, Ap \& Space Sci.  116, 401


\ref
 Aikawa, T. \& Sreenivasan, S. R. 1996, PASP 107, 238

\ref
 Alexander, D. R., Ferguson, J. W. 1994, ApJ 437, 879

\ref
 Benade, A. H. 1977, in {\sl Musical Acoustics}, Ed. E.L. Kent (Dowden,
Hutchinson \& Ross: Stroudsburg, PA), p. 110

\ref
 Buchler, J.R. 1997, in "Astrophysical Fallouts From Microlensing Projects",
Eds. R. Ferlet \,J.P. Maillart \& B. Raban, Editions Fronti\`eres, Gif-s-Yvette

\ref
 Buchler, J. R. \& Kov\'acs, G. 1986, {\it ApJ}, 303, 749 

\ref
 Cox J. P. 1980, {\sl Stellar Pulsations}, Princeton Univ. Press

\ref
 Cox, J. P. \& Giuli R. T. 1969, {\sl Principles of Stellar Structure} (New
York: Gordon and Breach)

\ref
 Cox, J. P. \& Guzik, J. A.  1996, in ``Luminous Blue Variables: Massive 
  Stars in Transition,'' Eds. A. Nota \& H. Lamers

\ref
 Cox, J. P., King, D. S., Cox, A. N., Wheeler, J. C., Hansen, C. J. \& Hodson,
S.W. 1980, Space Sci. Rev. 27, 529

\ref
 Deubner, F.-L. \& Gough, D. 1984, Ann. Rev. Astr. Astrophys. 22, 593 

\ref
 Eisner, E. 1967, J. Acoust. Soc. Am. 41, 1126--1146 

\ref
 Fraley, G. S. 1968, Ap\&SS, 2, 96

\ref
 Gautschy, A. \& Glatzel, W. 1990, MNRAS 245, 597

\ref
 Glasner, A. \& Buchler, J. R. 1993, {\sl A\&A}, 227, 69

\ref
 Glatzel, W. 1994, MNRAS 271, 66

\ref
 Iglesias, C.A. \& Rogers, F. J., 1996, {\sl ApJ}, 464, 943

\ref 
 Ledoux, P. \& Walraven, T. 1958, in {\sl Handbuch der Physik}, ed. S.
Fl\"ugge (Berlin : Springer), 51, 353
 
\ref
 Morse, P.M. \& Ingard, K.U., 1968, {\sl Theoretical Acoustics}, 
McGraw-Hill : New York)

\ref
 Morse P.M. \& Feshbach, H. 1953, {\sl Methods of Theoretical 
Physics}, (McGraw Hill, NY)

\ref Papaloizou, J.C.B., Alberts, F., Pringle, J.E., Savonije, G.J. 1997, 
MNRAS 284, 821--829

\ref
 Pesnell, W.D. \& Buchler, J.R.  1986, ApJ. 303, 740

\ref
 Saio, H., Wheeler, C.J.  \& Cox, J.P. 1984, ApJ 281, 318

\ref
 Unno, W., Osaki, Y., Ando, H., Saio, H., \& Shibahashi, H. 1989, {\sl
Nonradial Oscillations of Stars}, Univ. of Tokyo Press

\ref
 Zalewski, J. 1992, PASJ 44, 27

\end\bye